\documentclass[fleqn,usenatbib]{mnras}

\usepackage{newtxtext,newtxmath}

\usepackage[T1]{fontenc}

\DeclareRobustCommand{\VAN}[3]{#2}
\let\VANthebibliography\thebibliography
\def\thebibliography{\DeclareRobustCommand{\VAN}[3]{##3}\VANthebibliography}

\usepackage{graphicx}	
\usepackage{amsmath}	
\usepackage{ulem}
\usepackage{caption}
\usepackage{subcaption}
\usepackage{cancel}
\usepackage[left]{lineno}

\usepackage{xcolor}

\def\rr1#1{#1}

\newcommand{\rdmp}{\textsc{redMaPPer}~}
\newcommand{\metacal}{\textsc{metacalibration}~}

\title[Optical Signatures of Galaxy Cluster Triaxiality]{Towards Quantifying the Impact of Triaxiality on Optical Signatures of Galaxy Clusters: Weak Lensing and Galaxy Distributions}

\author[Shenming Fu, Yuanyuan Zhang, Camille Avestruz, Ruben Coronel]{
Shenming Fu,$^{1}$\thanks{E-mail: shenming.fu@noirlab.edu}
Yuanyuan Zhang,$^{1}$
Camille Avestruz,$^{2,3}$
Ruben Coronel$^{2,4}$
\\
$^{1}$NSF’s National Optical-Infrared Astronomy Research Laboratory, 950 North Cherry Avenue, Tucson, AZ 85719, USA\\
$^{2}$Department of Physics, University of Michigan, Ann Arbor, MI 48109, USA\\
$^{3}$Leinweber Center for Theoretical Physics, University of Michigan, Ann Arbor, MI 48109, USA\\
$^{4}$Department of Physics, Stanford University, 382 Via Pueblo Mall, Stanford, CA 94305, USA
}

\date{Accepted XXX. Received YYY; in original form ZZZ}

\pubyear{2024}

\begin{document}
\label{firstpage}
\pagerange{\pageref{firstpage}--\pageref{lastpage}}
\maketitle

\begin{abstract}
We present observational evidence of the impact of triaxiality on radial profiles that extend to 40 Mpc from galaxy cluster centres in optical measurements.   
We perform a stacked profile analysis from a sample of thousands of nearly relaxed galaxy clusters from public data releases of the Dark Energy Survey (DES) and the Dark Energy Camera Legacy Survey (DECaLS). Using the central galaxy elliptical orientation angle as a proxy for galaxy cluster orientation, we measure cluster weak lensing and excess galaxy density axis-aligned profiles, extracted along the central galaxy's major or minor axes on the plane-of-the-sky. Our measurements show a $\gtrsim2-3\sigma$ difference per radial bin between the normalized axis-aligned profiles.  The profile difference between each axis-aligned profile and the azimuthally averaged profile ($\sim\pm10-20\%$ along major/minor axis) appears inside the clusters ($\sim0.4$ Mpc) and extends to the large-scale structure regime ($\sim10-20$ Mpc). 
The magnitude of the difference appears to be relatively insensitive to cluster richness and redshift, and extends  further out in the weak lensing surface mass density than in the galaxy overdensity.  
Looking forward, this measurement can easily be applied to other observational or simulation datasets and can inform the systematics in cluster mass modeling related to triaxiality. 
We expect imminent upcoming wide-area deep surveys, such as the Vera C. Rubin Observatory's Legacy Survey of Space and Time (LSST), to improve our quantification of
optical signatures of cluster triaxiality.


\end{abstract}




\section{Introduction}

Galaxy clusters are the most massive gravitationally collapsed structures in our universe, providing both a probe for cosmology and a unique environment for galaxy formation \citep{allen11,kravtsovandborgani2012}.  Clusters form late in cosmic history; both relatively smooth mass accretion and mergers of galaxies and groups govern \rr1{their} mass growth.  Each galaxy cluster's mass accretion history, in turn, depends on the large scale structure and environment in which it resides.  The abundance and evolution of galaxy clusters is sensitive to the nature of dark energy \citep{vikhlinin09}.  But, one of the primary roadblocks to cluster-based cosmology lies in our ability to precisely and accurately meausure cluster masses.  With the advent of stage-4 cosmology surveys \citep{dodelson16}, the community is well-poised to utilize multi-wavelength information and Bayesian approaches towards improved precision and accuracy \citep{mulroyfarahi2019}. In fact, the weak gravitational lensing observable from optical surveys is now a standard for calibrating `observable-mass' relations \citep{mantz15,bocquet19}.

While most mass estimates that use galaxy cluster mass proxies assume azimuthally symmetric signatures and on-average spherical distribution, the underlying matter distribution of galaxy clusters more closely follow a triaxial shape \citep{kasunandevrard05,knebeandwiessner2006,despali2014}.  Cluster triaxiality is one of the known systematics to weak lensing mass estimates \rr1{\citep{becker2011,herbonnet19,zzhang2023}}.  Galaxy clusters whose major axis aligns with our line of sight will have boosted lensing signatures and systematically higher inferred masses, while those whose minor axis aligns with our line of sight will have suppressed lensing signatures \citep{osato2018,herbonnet22}.  The relative alignment of a galaxy cluster with our line of sight also affects other projected quantities, such as the \rr1{measured number of member galaxies}~\citep[i.e. the cluster `richness';][]{wu2022}. As a result, the imprints of cluster triaxiality on observations are correlated,  leading to a covariance between cluster observables\rr1{~\citep{zzhang2023}}.   
For example, clusters selected by an observable such as richness may also display a biased lensing signature \citep{2012MNRAS.426.1829N, 2022MNRAS.511L..30Z, 2022ApJ...933...48F}; cosmological studies need to account for such  selection effects. 
On the other hand, the impact of triaxiality on both the underlying mass and the galaxy distributions also allows for methods that constrain the angle and the ellipticity of the host dark matter halo ~\citep{shin2018,gonzalez21}.
The triaxial shape of galaxy clusters affects other signatures of galaxy clusters, including X-ray isophotes~\citep{rasia2013,mantz2015spa}, and can trace aspects of the galaxy cluster environment, such as filament alignment \citep{tempel2015,sifon2015,Gouin2020,lokken2022}.  

The shape and orientation of the central galaxy (CG) or the brightest cluster galaxy (BCG) can serve as proxies that describe its host cluster triaxiality.  Compared with measurements of the member galaxy distribution and lensing mass distribution, measurements of the CG are relatively less affected by noise. Additionally, CG measurements are more readily available for larger optical samples of galaxy clusters.  Other proxies, such as X-ray isophotes that trace the gas morphology, require many photon counts to constrain gas shapes and are more limited in availability~\citep[e.g.][]{rasia2013,mantz2015spa}. Similarly, the resolution of Sunyaev–Zel'dovich (SZ) effect observations hinders morphology measurements at high redshifts, but have been achieved for a small sample of clusters at relatively  low redshifts~\citep[e.g.][]{donahue16}. Such shape observations of higher redshift clusters exist for smaller samples~\citep[e.g.][]{rg2017,kitayama2023}.
Overall, existing optical searches provide an ample balance between coverage and depth to perform studies on cluster triaxiality in a large statistical sample~\citep{clampitt2016,shin2018}. 
CG shapes have already been used to quantify potential biases in weak lensing cluster mass estimates due to viewing angle~\citep{herbonnet19}, and CG orientations indicate the co-evolution with the intracluster medium~\citep{yuanwen2022}.  Simulations have identified measurable correlation between the CG orientation and the underlying cluster halo mass distribution~\citep{herbonnet22}. This, and other proxies of triaxiality are often used to identify physical mechanisms that tie galaxy clusters' observables and their mass accretion histories~\citep{chen19,machado21,deluca2021,sereno2021,mendoza2023,vallesperez2023}.  Observations have also provided concrete evidence of correlation in the alignment of various morphology measures, including that of the CG, gas, and underlying mass distribution, thereby motivating such proxies and further studies to understand correlated alignments~\citep{donahue16}.

The goal of this paper is to observationally investigate how cluster triaxiality impacts the distribution of the density of cluster member galaxies and galaxy cluster weak lensing signatures, and how these signatures co-vary with each other on different distance scales. 
In this work, we use the  (observed 2D-projected) orientation of the CG as a proxy for the orientation of the underlying galaxy cluster triaxial shape and as a tracer of triaxial alignments between cluster components, which is a more commonly available measurement for optical cluster samples, such as those used in this work -- the recent Dark Energy Survey (DES) Y3 catalogue~\citep{gatti21} and the Dark Energy Spectroscopic Instrument (DESI) Legacy Image Surveys (hereafter Legacy Surveys) catalogue~\citep{dey19}.   
We use the orientation angles to align our stacked measurements of cluster member galaxy distributions and weak lensing signatures and to separate these measurements along the cluster major and minor axes.   
We note that, our approach is complementary to the analysis done by~\cite{shin2018}, where they examined the relationship between the CG, cluster member galaxy distribution, and underlying mass distribution traced by weak gravitational lensing with quadrupole measurements.  We additionally discuss how the optical signatures connect the triaxial alignment and the surrounding large scale environment, motivated by the theoretical analysis done by~\cite{osato2018}.  

We organize the paper as follows.  In Section~\ref{sec:theory}, we briefly describe the theoretical underpinnings of using the weak gravitational lensing signature of galaxy clusters to trace their underlying mass distribution, which is largely triaxial.  
In Section~\ref{sec:datasets}, we describe the galaxy cluster samples and associated weak lensing shape measurement catalogues, photometric redshifts (photo-zs) and photometry that we use in this study.  Section~\ref{sec:results} lays out the results, quantifying how radial profiles of both the galaxy number density (in and around galaxy clusters) and the weak lensing signature differ when measured along the directions of CG major and minor axes.  We also discuss potential effects due to richness, redshift, or red-sequence selection in our sample.
Finally, we provide our discussion of the results in Section~\ref{sec:discussion}, including considerations surrounding robustness of our measurements, methods of CG orientation \rr1{determination}, correlation between lensing signature and number density.
We summarize our work and talk about possible future directions in Section \ref{sec:summary}.
In the Appendix, we show some extra tests and details, and a flow chart for our pipeline.

\section{Theory}
\label{sec:theory}
The mass of a foreground galaxy cluster deflects the light from distant background galaxies via gravitational lensing. 
The deflection of a light source on the image plane can be described by 
\begin{equation}
\boldsymbol{\alpha}=\boldsymbol{\theta}-\boldsymbol{\beta}~,     
\end{equation}
where $\boldsymbol{\theta}$ and $\boldsymbol{\beta}$ denote the observed position and the original position of the source respectively. 
The deflection and the cluster surface mass distribution $\Sigma(\boldsymbol{\theta})$ are connected through a potential, so that 
\begin{equation}
\boldsymbol{\alpha}=\nabla\phi~.    
\end{equation}
Here, 
\begin{equation}
\phi(\boldsymbol{\theta})=\frac{1}{\pi}\int \kappa(\boldsymbol{\theta'}) \ln{|\boldsymbol{\theta}-\boldsymbol{\theta'}|}\,\mathrm{d}^2\theta'~.
\end{equation}
The \textit{convergence} $\kappa$ 
satisfies 
\begin{equation}
\nabla^2\phi=\nabla\cdot\boldsymbol{\alpha}=2\kappa,~\kappa(\boldsymbol{\theta})=\frac{\Sigma(\boldsymbol{\theta})}{\Sigma_\textrm{crit}}~.
\end{equation}
$\Sigma_\textrm{crit}$ is the critical surface mass density satisfying
\begin{equation}
    \Sigma_\textrm{crit}=\frac{v_\textrm{c}^2}{4\pi G}\frac{D_\textrm{s}}{D_\textrm{l} D_\textrm{ls}}~,
\end{equation}
where $D_\textrm{s}$, $D_\textrm{l}$, and $D_\textrm{ls}$ are the angular diameter distances between the source and the observer, the lens and the observer, the lens and the source respectively; $v_\textrm{c}$ is the speed of light.

The differential lensing effect (the Jacobian matrix) is
\begin{equation}
\frac{\partial \boldsymbol{\beta}}{\partial\boldsymbol{\theta}}=\delta_{ij}-
\phi_{,ij}
=(1-\kappa) I-\gamma_1
\begin{pmatrix}
1 & \\ 
  & -1
\end{pmatrix}
-\gamma_2
\begin{pmatrix}
   & 1\\ 
 1 & 
\end{pmatrix}
~,
\end{equation}
where $i,j=1,2$ and the \textit{shear} components 
\begin{equation}
\gamma_1=(\phi_{,11}-\phi_{,22})/2,~ \gamma_2=\phi_{,12}=\phi_{,21}~, 
\end{equation}
and thus the convergence gives an isotropic magnification, while the shear produces distortions along two axes separated by 45 degrees~\citep{narayan1996,bartelmann2001}.

\begin{figure*}
    \includegraphics[width=1.95\columnwidth]{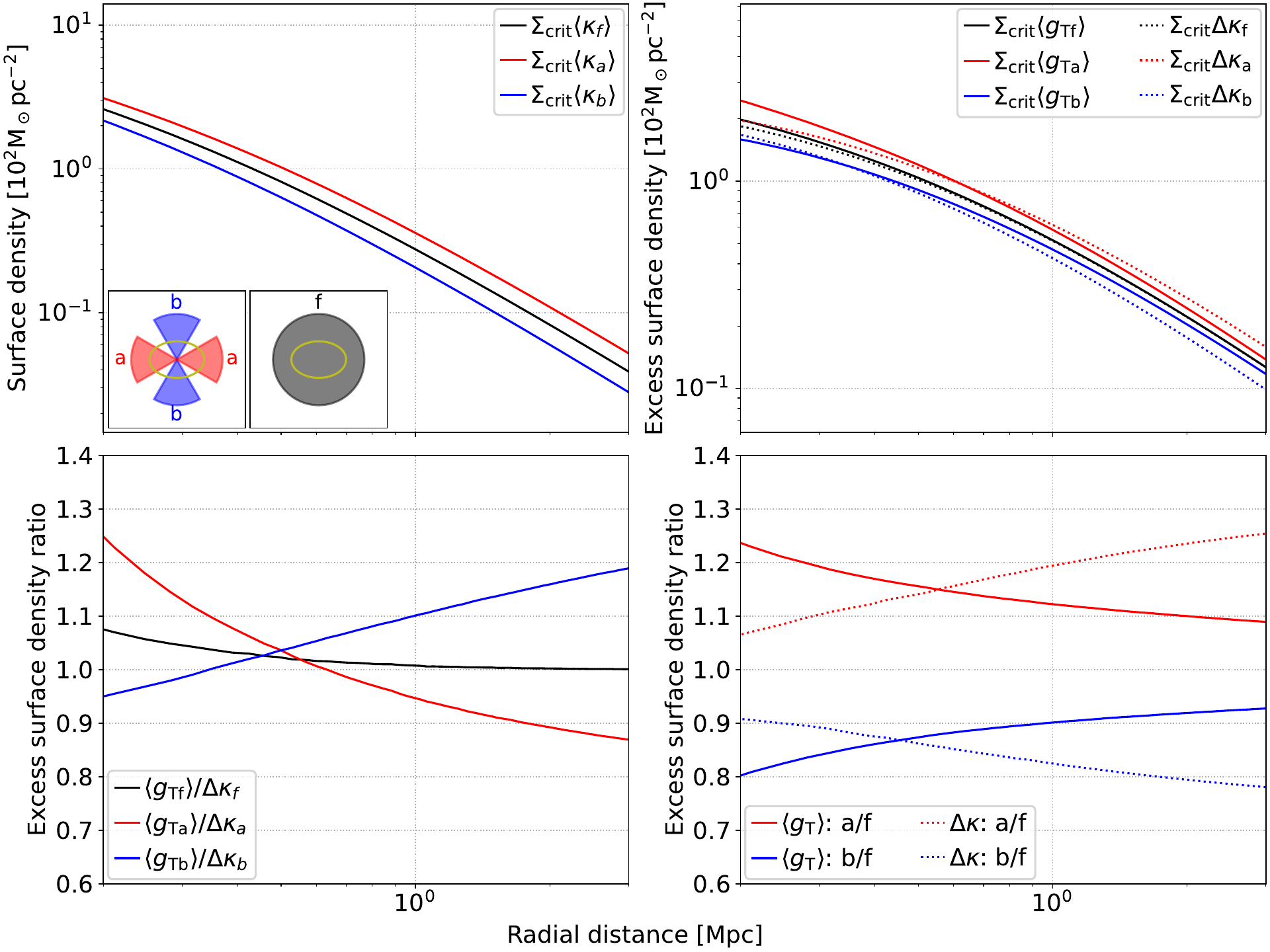}
    \caption{
    Example of lensing effects of an elliptical NFW cluster halo. We consider a cluster that is at redshift 0.47 and has \rdmp richness 33 (median of the observational sample studied later in the paper; Section \ref{sec:catalog_stacking}), and convert that to $M_\textrm{200c}=2\times10^{14}M_\odot$, based on the DES Y1 mass-richness relation~\citep{mcclintock19} and typical $M_\textrm{200m}/M_\textrm{200c}$ ratio\rr1{~\citep{Diemer2018,kovacs2022}}. The subscript 200 means the average density of the spherically  enclosed overdensity  mass is 200 times the critical (c) or mean matter (m) density of the universe at that redshift. 
    \rr1{\citet{mcclintock19} uses a slightly different cosmology (flat $\Lambda$CDM $H_0=70$~km~s$^{-1}$~Mpc$^{-1}$ and $\Omega_\textrm{m}=0.3$), but we do not expect it to strongly affect the demonstration of triaxiality signatures using mock data here. }
    We use the concentration-mass relation and cosmological parameters  described by~\citet{child2018}.
    We compute the lensing convergence and shear using the method described by~\citet{oguri2010b} \rr1{and}~\citet{oguri2010a}, so that the surface density has an elliptical form \rr1{$\kappa(\boldsymbol{\theta})=\kappa_\textrm{NFW}(\xi=\sqrt{q\theta_1^2+\theta_2^2/q})$}, where NFW is the spherical profile described by~\citet{navarro96}, and $q$ is the axis ratio $b/a$ (here we select 2/3 to be consistent with our results in Section \ref{sec:results}). We consider the radial distance range for weak lensing from 0.2 to 3~Mpc ($r_\textrm{200c}$ is 1~Mpc). The correlated large-scale structure (LSS; the two-halo term) is not modelled and can be aligned with the halo, and is beyond the scope of this demonstration. 
    We select two samples ($a, b$) of areas near the major and minor axes of the ellipse ($\pm 30 \deg$) and compute the respective \textit{averages} along the arcs given the radial distance, and we use the azimuthal \textit{average} over all position angles as \rr1{a} fiducial sample ($f$), similar to our strategy for processing the real data (Section \ref{sec:CG_angle}). After averaging, the results are unrelated to the directions of the X-axis and Y-axis.
    \textit{Top left}: Surface density comparison ($a>f>b$). The inset plots show the angle ranges of the three samples and the shape of the elliptical halo, which are the same in other subplots.
    \textit{Top right}: Comparisons of excess surface density for different samples (\rr1{$\Delta\Sigma_i=\Sigma_{\rm crit}\Delta\kappa_i$ and $\Delta\Sigma_a>\Delta\Sigma_f>\Delta\Sigma_b$}), and \rr1{goodness of the approximation shown in Eq.~\ref{eq:delta_Sigma}, $\Delta\Sigma_i\sim\Sigma_{\rm crit}\langle g_{ {\rm T}i}\rangle$, $i=f,a,b$. }
    \textit{Bottom left}: Ratios between reduced shear and excess convergence, which shows how accurate the excess density estimate is (error \rr1{$\lesssim25\%$}).
    \textit{Bottom right}: Ratios between the major-axis sample $a$ (or the minor-axis sample $b$) and the fiducial sample ($f$) for the reduced shear and the excess density. \rr1{Later in Section \ref{sec:results}, we use lensing effects and galaxy photometry in the real observational data to study those ratios respectively.} The ratios along the two directions are nearly symmetric with respect to 1.
    }
    \label{fig:elliptical_halo_model}
\end{figure*}

Using a loop around the cluster on the image plane and 2D divergence theorem, we have 
\begin{equation}
\iint 2\kappa \, \mathrm{d}^2\theta = 
\oint (\nabla\phi \cdot \hat{n}) \, \mathrm{d}l = \int \alpha_n \, \mathrm{d}l~,
\end{equation} 
where $\alpha_n$ is the projection of $\boldsymbol{\alpha}$ along the outward normal direction of the loop. 
Traditionally, a circular loop with radius $\theta$ is considered, 
then 
\begin{equation}
\langle\gamma_\textrm{T}\rangle|_{\theta}=\Bar{\kappa}(<\theta)-\langle\kappa\rangle|_{\theta}\equiv\Delta\kappa(\theta)~, 
\end{equation}
where $\langle\cdot\rangle|_{\theta}$ is the average at that radius, 
$\gamma_\textrm{T}$ is the shear component tangential to the loop,  
and $\Bar{\kappa}(<\theta)$ is the mean convergence inside that loop; 
this is true for an arbitrary $\kappa$ distribution~\citep{schneider2005}. $\Delta\kappa$ is the excess convergence.
If we consider a circular sector (pie-shaped) loop instead, i.e. two radii that cross at the mass centre and span an angle, with a circular arc, this equation still holds for an axis-symmetric lens.
Going further, for an elliptical lens, if this angle is small, or the mass distribution is nearly axis-symmetric (which is likely true for the nearly relaxed cluster sample studied in this paper), then $\alpha_n$ is small on the two straight lines, and the above equation still holds approximately (Figure \ref{fig:elliptical_halo_model}).

In the weak lensing regime, both $\kappa,~\boldsymbol{\gamma}\ll1$. 
The lensing observable is the \textit{reduced shear} 
\begin{equation}
\boldsymbol{g}=\frac{\boldsymbol{\gamma}}{1-\kappa}~, 
\end{equation}
and in weak lensing $\boldsymbol{g}\sim\boldsymbol{\gamma}$. 
The mean shape of background galaxies satistifies
\begin{equation}
    \langle \boldsymbol{g} \rangle\sim\langle\boldsymbol{\gamma}\rangle\sim \langle \mathbf{\mathsf{R}} \rangle^{-1} \langle \boldsymbol{e} \rangle~,
    \label{eq:shear_response}
\end{equation}
where the matrix $\mathbf{\mathsf{R}}$ is a per-object shear response (Section \ref{sec:shape_catalog}).
The excess surface mass density $\Delta \Sigma (R)$ satisfies
\begin{equation}
\langle g_\textrm{T} \Sigma_\textrm{crit} \rangle|_R \sim \langle \gamma_\textrm{T} \Sigma_\textrm{crit} \rangle|_R = \overline{\Sigma} (<R) - \langle\Sigma\rangle|_R \equiv \Delta \Sigma (R)~, 
\label{eq:delta_Sigma}
\end{equation}
and links the lensing observable to the mass distribution within a (2D) radial distance $R$ for a circular loop. 
For a circular sector area (Section \ref{sec:CG_angle} and Section \ref{sec:results}), we can similarly define an `effective excess surface density' ($\widetilde{\Delta\Sigma}$) using $\langle g_\textrm{T} \Sigma_\textrm{crit} \rangle$ from a sample of background galaxies inside. Also, since the distribution  of cluster galaxies approximately traces the dark matter (DM) halo, we can similarly define an excess number density, $\Delta n=\overline{n}(<R)-n(R)$, for comparison. 

The lensing effect of cluster triaxiality has been studied in simulations using elliptical NFW-like models~\citep{oguri2003}. Their projected isodensity contours are concentric, aligned, and elliptical with fixed ellipticity (homoeoidal).  
This motivates us to consider two profiles along two perpendicular directions on the projection plane (plane of the sky) -- the major and minor axes of the ellipse -- where the two profiles should have the largest difference. 
We use their difference to detect the halo ellipticity, though the halo does not necessarily need to have an NFW shape. 
In addition, the profile can describe either the mass distribution, which can be observed via lensing, or the number distribution of cluster galaxies, which traces the mass distribution (Figure \ref{fig:CG_angle_cut}). 
However, the goal of this paper is not to fit the data with an elliptical model to exactly measure the ellipticity, but rather to present a simple method that can analyse the lensing signature of cluster triaxiality without model assumption, and to study how it is related to other physical quantities. 
We demonstrate our method using public datasets.

\rr1{For the analysis of real observational data in this} paper, we use the same cosmological parameters as ~\citet[][]{child2018} \rr1{and~\citet[][]{Korytov2019}} -- flat $\Lambda$CDM with $H_0=71$~km~s$^{-1}$~Mpc$^{-1}$, $\Omega_\textrm{m}=0.2648$.  
We note that the physical radial distance \rr1{and the critical surface density} in each cluster \rr1{are the places} where the cosmological parameters get involved (through the angular diameter distance); switching between commonly accepted values of the cosmological parameters only changes the distance by \rr1{$\lesssim1\%$ and the critical surface density by $\lesssim2\%$}. 
We use physical units rather than comoving for radial distances. 
The magnitudes are in the AB system~\citep{oke1983}.


\section{Datasets and Methods}\label{sec:datasets}
\subsection{Cluster samples}\label{sec:cluster_sample}

The cluster sample used in this paper is identified by the \rdmp algorithm \citep{rykoff2014}, which is based on searching of cluster galaxy red-sequence features in photometric datasets. The \rdmp algorithm has been applied to a handful of galaxy catalogues including the Sloan Digital Sky Survey (SDSS) and DES, tested with multi-wavelength data from cosmic microwave background (CMB) and X-ray observations \citep{redmapper2, redmapper3} as well as spectroscopic follow-ups \citep{redmapper4}, and used in multiple galaxy cluster  abundance cosmology studies \citep{costanzi2019, costanzi21, abbott2020, park2023}.

We use the \rdmp cluster catalogue derived from the DES Year 1 photometric datasets, a publicly-accessible catalogue described by \citet{mcclintock19}\footnote{\url{https://des.ncsa.illinois.edu/releases/y1a1/key-catalogs/key-redmapper}}. 
For each galaxy cluster in the sample, \rdmp also identifies five CG candidates with centring probability assigned to each one (\texttt{P\_CEN}). We use the most probable CG from this list. Because the CG orientation angle is a crucial component in this analysis, we require the clusters to have a centring probability  of at least  80\%\footnote{$\texttt{P\_CEN}=0.8$ is roughly the mean centring probability of \rdmp clusters when matched with SZ/X-ray observations~\citep{rykoff2016}.}  associated with the most probable CG ($\max\{ \texttt{P\_CEN}\}\geq0.8$). 
\rr1{
In addition, the higher values of \texttt{P\_CEN} indicate the lower probability for other galaxies being ``central''-like, in which case the cluster is less likely to have multiple central-like galaxies. Since multiple central galaxies suggest a recent merger history~\citep[e.g.][]{Edwards2012,Mann2012,Furnell2018}, 
we expect the selected clusters with high \texttt{P\_CEN} to be relatively more relaxed. 
} 
 
We also use the cluster redshift information available in the catalogue (\texttt{Z\_LAMBDA}), a photometric redshift calculation estimated from the cluster red sequence features calibrated with archival spectroscopic redshifts (spec-zs). This cluster galaxy red sequence redshift has been shown to be nearly unbiased, with a percentage level redshift scatter ($\sigma_z$): the median of $\sigma_z/(1+z)\simeq 0.006$~\citep{mcclintock19}, sufficient for this analysis. After the selection on \texttt{P\_CEN}, the remaining 4300 clusters span richness $20<\lambda<235$ and redshift $0.2<z<0.86$. We note that the clusters that are above redshift 0.65 might have a completeness issue~\citep{mcclintock19}. However, those clusters only cover a small fraction ($3\%$) and our results are not strongly sensitive to the redshift. Therefore, we still include those clusters to reduce the statistical noise. 

Multi-wavelength studies of the \rdmp clusters have provided further insight into the catalogue's performance. X-ray \citep{zhang2019_centering} and CMB \citep{bleem_2020} studies confirmed that the most probable CGs of the \rdmp clusters (regardless of their \rdmp assigned centring probabilities) are correct around 75\% of the time, assuring us the quality of the CG selection; the cut on \texttt{P\_CEN} further improves the selection. 
\rr1{The coordinates of the most probable CG are taken as the cluster coordinates. }
In this paper, we also rely on using the richness quantity in the catalogue as a mass proxy. X-ray and weak lensing \citep{mcclintock19} studies show that the galaxy clusters with richness above 20  correspond to a mass range of roughly  $10^{14} \mathrm{M}_\odot h^{-1}$ and above.

Finally, to qualitatively validate the results of this paper, we also test the analysis with a cluster catalogue derived from the Legacy Surveys (LS) imaging data  and an independently developed cluster finding algorithm \citep{zou21}. \cite{zou21}  searched for clusters based on galaxy overdensity and photometric redshifts. The final catalogue also includes a BCG selection, the brightest galaxy in the $r$-band within 0.5 Mpc of the galaxy overdensity peak, which we use for comparison analyses. 
\rr1{Similar to the \rdmp cluster sample, we use the BCG coordinates as the cluster coordinates. 
In relaxed clusters, the separation between the BCG and the cluster centroid is expected to be small~\citep[e.g.][]{Zenteno2020}. 
Because we lack sufficient X-ray and spec-z data for this cluster dataset, we use the reported galaxy density peak as  the cluster centroid. } 
\rr1{We choose clusters that have a small distance between the BCG and the density peak ($< 0.3$ Mpc),  
and we find that using an even smaller cut ($0.1$ Mpc) gives similar cluster triaxiality signatures but with larger error bars. }
We select clusters with redshifts between 0.2 and 0.65 \rr1{(false detection rate $<5\%$)}, roughly within the DES footprint, and with a luminosity richness larger than 30 and a member galaxy count larger than 20. 
\rr1{Here, the luminosity richness corresponds to the total member galaxy luminosity within 1~Mpc in units of a characteristic luminosity defined in \citet{zou21}.  A luminosity richness of 30 roughly corresponds to a \rdmp richness of 20. With this luminosity richness threshold, we hope to select a cluster sample similar in mass to the \rdmp cluster sample.  }
We show the results derived from the LS clusters in Appendix \ref{sec:ls_cluster}. 
\rr1{In the future, the large data sets coming from the DESI spec-z survey and the eROSITA X-ray survey can be used to better determine the cluster dynamical state~\citep[e.g.][]{mantz2015spa,Cui2017,deluca2021}. }

\subsection{Shape catalogues}\label{sec:shape_catalog}

We use the publicly available DES Y3 shape catalogue~\citep{gatti21}\footnote{\url{https://des.ncsa.illinois.edu/releases/y3a2/Y3key-catalogs}} for the cluster weak lensing measurements, which was derived using the  \metacal~\citep{huffandmandelbaum17,sheldonandhuff17} algorithm and the \texttt{ngmix} package\footnote{\url{https://github. com/esheldon/ngmix}} from the first three years of the DES imaging observation. The shear measurement~\citep{zuntz18} was derived from reduced and calibrated multi-epoch and multi-band images for each object and self-calibrated on those images. Note that this shape catalogue is based on the Year-1 to Year-3 observations of DES, rather than the Year-1-only observation, which was the basis of the cluster catalogue. A DES Year-1-only shape catalogue also exists~\citep{zuntz18}, but this Y3 version features a few improvements, including PSF modeling using the PIFF package~\citep{jarvis2021} and a more uniform sky coverage. 
Information about this shape catalogue is listed in Table~\ref{tab:catalogs}.

To determine the redshifts of the lensing source galaxies, we use the Directional Neighbourhood Fitting (DNF) photometric redshifts \citep{devicente16} available in the DES Y3 data release. 
DNF finds the nearest neighbours of a galaxy in the multi-magnitude space to determine the photo-z based on a spec-z training set and galaxy fluxes and colours.
We use the \texttt{ZMEAN\_SOF} in the catalogue, the best DNF  redshift estimate, for individual source galaxies. 
Those redshifts are validated against the photometric redshifts of galaxies in the COSMOS field using the method outlined by \citet{hoyle18} for the DES Y1 data. 

To test the robustness of our results, we also analyse the Legacy Surveys (LS)\footnote{\url{https://www.legacysurvey.org}} DR9 shape catalogue~\citep{dey19} and photometric redshift catalogue~\citep{zhou21}. 
This information is also listed in Table~\ref{tab:catalogs}. 
The Legacy Surveys shape catalogue is derived using the Tractor algorithm \citep{2016ascl.soft04008L} by parametric light profile model fitting. 
Those shape measurements have been used to derive accurate cluster mass estimates when calibrated against the CFHT Stripe 82 (CS82) catalogue~\citep{phriksee2020}. In this analysis, we do not apply those additional calibrations as we are analysing the \textit{relative} amplitudes of the cluster lensing signals along different directions. 
The photometric redshift catalogue is derived using a random forest method \citep{breiman2001random}.
We discuss the details in Section \ref{sec:discussion}.


In the \metacal algorithm, the mean shear of an ensemble of sources can be estimated by the product of the inverse of their mean shear response and mean shape (Eq. \ref{eq:shear_response}).
The shear response matrix $\mathbf{\mathsf{R}}$ for each source is estimated by applying a small artificial shear to the image and computing the derivative of the measured ellipticity as a function of the applied shear.
For a sample that includes a selection, 
the selection can change the mean shape and bias the mean shear response. Therefore, a selection response, which is based on the selection on the catalogues
of the artificially sheared images, needs to be considered as well, and it is usually at the percent level of the shear response.  
The \metacal selection information has been included in the DES Y3 catalogue.

For each cluster, we search for galaxies that are within an angle corresponding to 40~Mpc at the cluster's redshift (\texttt{Z\_LAMBDA}).
The DES Y3 shape catalogue contains coordinates and is given \rr1{over} the whole DES footprint. 
To save computational resources and time, instead of searching the whole catalogue for source galaxies associated with each cluster, we first divide the DES Y3 catalogue into HEALPix pixels~\citep{2005ApJ...622..759G,Zonca2019}\footnote{\url{http://healpix.sf.net/};~\url{https://github.com/healpy/healpy/tree/main}}. We use $\texttt{NSIDE} = 32$ and RING ordering ($\sim3\deg^2$ per pixel) for the HEALPix division.
In those pixel catalogues, we record the \metacal quantities (per-galaxy ID, shape, response, and weight) and the selection information for both sheared and unsheared versions, and also the DNF redshifts.
Then we search for pixels around one cluster, and combine the pixel catalogues associated with that cluster.

Next, we select galaxies which have redshift \rr1{values of} at least 0.1 beyond the cluster's redshift ($\texttt{ZMEAN\_SOF}>0.1+\texttt{Z\_LAMBDA}$). This could potentially cause an extra selection response. However, the DNF redshifts measured on the sheared images are not available; the computation of that selection response is beyond the scope of this paper. We expect its effect on our results to be negligible, especially on the relative amplitude of \rr1{the cluster} lensing signal, as the artificial shear will not affect the photometry strongly. We also make a selection of $\texttt{ZMEAN\_SOF}<1.5$ to remove high-redshift sources because they are possibly not well detected or measured, although their fraction over the whole footprint is only $\sim7\times10^{-5}$ after using the \metacal selection. 

We then compute the mean response of this cluster field by combining the mean shear response and the mean selection response: $\langle\mathbf{\mathsf{R}}\rangle=\langle\mathbf{\mathsf{R}}_\gamma\rangle+\langle\mathbf{\mathsf{R}}_\textrm{s}\rangle$.
Here, the shear response of each source is given in the Y3 catalogue, and we compute the mean of the sources selected for the \metacal weak lensing analysis. 
The mean selection response is the ratio between the difference of the mean (unsheared) ellipticity under the \metacal selections on the sheared images and the difference of the applied shear.
Finally, we use the average of the diagonal terms (which are usually close) of $\langle\mathbf{\mathsf{R}}\rangle$ as a single response value ($\mathcal{R}$) to reduce the noise; the off-diagonal terms are relatively small and usually at the percent level or lower~\citep[e.g.][]{mcclintock19}. 
We use this $\mathcal{R}$ value for the whole cluster field as we treat each cluster independently. This allows us to account for the variation over the DES footprint, and each cluster field is large enough (a few degrees in diameter) for statistics.

After the above steps, we compute the per-object reduced shear estimates by $\hat{g}_i=e_i/\mathcal{R}$, where $i=1,2$,  
for each source selected for the \metacal weak lensing analysis. We record them and the coordinate, redshift, and weight in a shape catalogue for the cluster.

Now we build an excess surface density profile using the shape catalogue.
We divide the cluster radial distance into 8 logarithmic bins from 0.2 Mpc to 40 Mpc. We skip the central region to reduce the effects of blending, strong lensing, and limited background sources.
We then compute the tangential and cross components of the reduce shear by Eq. \ref{eq:g_t}, where $\varphi$ is the position angle towards the cluster centre (top North and left East),  as the flat sky approximation is still valid, and we compute per-object excess surface density estimates $\Delta \hat{\Sigma}_i = \hat{g}_i \Sigma_\textrm{crit}$, where $i=\textrm{T,~X}$.

\begin{equation}
    \hat{g}_\textrm{T}=- \hat{g}_1 \cos(2 \varphi)- \hat{g}_2 \sin(2 \varphi);~
    \hat{g}_\textrm{X}=\hat{g}_1 \sin(2 \varphi) - \hat{g}_2 \cos(2 \varphi)~.
    \label{eq:g_t}
\end{equation}

In each radial bin, if the number of sources is larger than 1, we compute their $\langle \hat{g}_i \Sigma_\textrm{crit} \rangle  \sim  \langle {g}_i \Sigma_\textrm{crit} \rangle$ and its error bar by bootstrap resampling; otherwise we set the result to be NaN (not a number).
In the resampling, we randomly pick half of the source sample in the bin with probabilities (weights) and replacement, and record the mean of $\Delta \hat{\Sigma}_i$. We repeat this 1000 times, and use their average to be $\langle \hat{g}_i \Sigma_\textrm{crit} \rangle$ and their standard deviation divided by $\sqrt{2}$ to be the error bar. 
We test and find \rr1{that} the direct weighted mean of all sources is close to the resampling mean.
Here the probability/weight is proportional to the product of the \metacal weight ($\sigma_\gamma^{-2} \sim \sigma_g^{-2}$) and $\Sigma_\textrm{crit}^{-2}$~\citep{sheldon2004}.
Using this weighting scheme decreases the error bars but does not strongly affect the results.

We note that there could be some extra multiplicative systematics e.g. a blending-related bias $\sim-2\%$~\citep{maccrann22}, and a boost factor~\rr1{\citep{varga2019}} which addresses the contamination of cluster galaxies; they both dilute the lensing signal. 
We expect those factors will generally be reduced in the relative amplitude of lensing signal, and thus we do not correct for them.
In fact, we find that even without the correction for those multiplicative systematics, along the CG major axis the stacked lensing signal is stronger than the one along the minor axis, but the blending effect and the boost factor would also be stronger along the major axis, since more cluster galaxies tend to reside there (Section \ref{sec:results}).

\rr1{When testing the results with the LS data products, we access the LS DR9 shape and photo-z catalogues through the NOIRLab Astro Data Lab\footnote{\url{https://datalab.noirlab.edu/ls/ls.php}}}.
For each cluster field, we make cone searches in \rr1{the} \texttt{ls\_dr9.tractor\_s} and \texttt{ls\_dr9.photo\_z} catalogues (matched by \texttt{ls\_id}) and select sources using these criteria: \rr1{their} signal-to-noise ratios (SNRs) are larger than 5 in $g,r,z$-bands; the mean and median of \rr1{their} photo-z PDF are \rr1{higher} than the cluster's redshift \rr1{value by} 0.1 \rr1{but lower than the redshift value of} 1.5; \rr1{their} (pre-PSF) morphological model \rr1{types are} not REX (round exponential galaxy, which is not ideal for lensing analysis); the half-light \rr1{radii of their models are} large than $0''$ (the PSF/stellar model type is a delta function/point source, and has a half-light radius of $0''$). We note that the model's ellipticity sign needs to be inverted so that the first component is along the X-axis (East to West) and the second \rr1{component} is 45 degrees above that anticlockwise. 
We then similarly build an excess surface density profile \rr1{of the cluster} using $\langle e_i \Sigma_\textrm{crit} \rangle  \sim  \langle {g}_i \Sigma_\textrm{crit} \rangle$ and estimate the \rr1{profile} error bars using bootstrap resampling. 
We note that adding weights based on the variance of the ellipticity given in the catalogue does not improve the \rr1{measurements} greatly, and we thus skip the weights here.


\begin{table*}
    \centering
    \begin{tabular}{lllll}
    \hline
    Datasets & Cluster catalogues & Shape catalogues & Photometry catalogues & Photometric redshift catalogues \\
    \hline
 
    Principal  & DES Y1 \rdmp & DES Y3   & DES DR2  
    & DNF for DES Y3 \\
    & \citep{mcclintock19} & \citep{gatti21}   & \citep{abbott21} & \citep{devicente16} \\
     & & & & \\
    Cross-checking & LS DR8 clusters & LS DR9  
    &  LS DR9  & LS DR9 photo-z (random forest) \\
    &\citep{zou21} & \citep{dey19} 
    &  --  & \citep{zhou21} \\
    \hline
    \end{tabular}
    \caption{Summary of the catalogues used in this work. The shape and photo-z catalogues are used for lensing analysis, while the photometry catalogues are used for number density analysis. 
    Our main results come from the principal dataset (\textit{top}; Section~\ref{sec:results}), and we test the robustness of our results by replacing some catalogue with the one in the cross-checking dataset (\textit{bottom}). For example, we test the lensing results using the LS shapes and photo-zs but keeping the cluster catalogue unchanged, and similarly we test the number density results using the LS photometry (Section~\ref{sec:robustness}); we also test our method using the clusters detected in LS DR8 but still using the DES shapes, photometry, and photo-zs (Appendix~\ref{sec:ls_cluster}). 
    }
    \label{tab:catalogs}
\end{table*}

\subsection{Photometry catalogues}\label{sec:photometry_catalog}
In this work, we also measure the galaxy density distribution inside the galaxy clusters. The galaxy catalogue used in this analysis is from the DES Data release 2 \citep[DR2;][]{abbott21}, which is a public release of the photometry information of the galaxies, stars, and other astronomical objects derived from the full six years of  DES observations. This is the deepest and widest DES data release, and contains significantly more data than the Year-1-only observation (which is used to construct the galaxy cluster catalogue). We use the DR2 (instead of DR1) catalogues because of their deeper depth and more uniform sky coverage, which both improve the precision of the galaxy density calculation.  

The DR2 photometry catalogues are derived by the DES data management (DESDM) team \citep{sevilla2011, morganson2018} based on the coadded Year 1 to Year 6 imaging data. The object detection is performed using the \texttt{Source Extractor} software \citep{bertin1996}, while the photometric measurements are performed with both the \texttt{Source Extractor} software and the \texttt{ngmix} algorithm -- a Gaussian mixture fitting algorithm applied to multi-epoch, multi-band imaging data. To compute the galaxy overdensities in the clusters, 
we select all photometric objects with SNR above 5, \texttt{flags} below 4 (well-behaved objects), \texttt{imaflags\_iso} being 0 (no missing/flagged pixels in the source in all single epoch images)  in all bands ($g,r,i,z,Y$), and $\texttt{extended\_class\_coadd}$ above 1 (mostly and high-confidence galaxies) in the DR2 main catalogue; the two quality flags cuts are recommended by DR2 and then the morphological object classification selection produces a benchmark galaxy sample~\citep{abbott21}.
Because the DES DR2 photometry catalogue has 95\% completeness magnitude limits of 24.3, 24.0, 23.7 in $r,i,z$-bands respectively  for objects matched and measured with SNR above 10 in the HSC-PDR2 Deep+UltraDeep fields~\citep{aihara2019}, we also add cuts at those magnitudes to ensure a uniform galaxy selection around each cluster. The $r,i,z$ are the bands where DESDM detects objects, and where we use colours to obtain a red-sequence sample (Section~\ref{sec:red-sequence}); we note that $g-r$ has a larger scatter than $r-i$ and $i-z$, and the $Y$-band depth is shallower.

Similarly, to check the robustness of the results, we also make use of the Legacy Surveys DR9 photometry catalogues \citep{dey19}. Those catalogues are derived using the Tractor software \citep{2016ascl.soft04008L} from the images collected by the Dark Energy Camera Legacy Survey (DECaLS) as well as archival images, and we use the 9th data release. Those catalogues generally reach 5-sigma point source depths of 24.8, 24.2, 23.3 magnitude in $g,r,z$-bands (and $\sim0.6$ mag deeper in the DES footprint), and have been designed to \rr1{achieve} great uniformity through their covered footprint. We select objects that have SNRs above 5 in $g,r,z$ and model half-light radius above 0" (i.e. all model types except PSF/stellar). 
Though LS DR9 did not report 95\% completeness magnitude limits, 
we find that adding a cut at $\sim23$th magnitude  does not greatly affect the relative amplitude of excess number density.
Information about those photometry catalogs are also listed in Table \ref{tab:catalogs}.

For each cluster, we make queries through the Astro Data Lab\footnote{\url{https://datalab.noirlab.edu/des/index.php}} 
and use 40 Mpc as the radial distance limit.
We apply the above cuts, and then derive a galaxy number density profile for the cluster using the above 8 bins and an extra bin: $[0,0.2)$ Mpc. 
We also compute an \textit{excess} number density profile [$\Delta n(R)\equiv\Bar{n}(<R)- n(R)$], so that the value in each bin is the difference between the density of all previous bins (smaller radii) and the density of the current bin; the excess number density in the first bin is set to be NaN.
After averaging a large ensemble of clusters, the foreground/background galaxy density will be removed in the excess number density automatically.
Near the cluster centre, the lensing deflection can potentially affect the background galaxy distribution, but we expect this effect to be small in most radial distance bins, especially in the relative amplitude.
Additionally, we find that averaging the excess number density profiles gives almost the same results as averaging the number density profiles and then computing excess number density from the mean number density profile. We present the details of the stacking process in Section~\ref{sec:catalog_stacking}.


\subsection{CG angle measurement and galaxy sample selection}\label{sec:CG_angle}

In order to quantify galaxy cluster triaxiality in the observational datasets, we consider the mass and galaxy distributions along the major and minor axes of the elliptical 2D projection.
The orientations of the cluster halo and member galaxy distribution can be inferred from lensing signal and photometry respectively, which can be noisy in individual clusters.
On the other hand, the orientation of the CG approximately traces that of the cluster and is relatively well-defined and easier to measure. 
Since the CG is \rr1{generally} much brighter and larger than other member galaxies, the effect of blending is also small. 

To locate the CG in each \rdmp cluster selected by $\texttt{P\_CEN}$ (Section~\ref{sec:cluster_sample}), we make a small cone search ($0.01\deg$ in radius) near the cluster centre in the DES Y3 catalogue without selecting the redshifts and computing the responses.
We then associate this catalogue with the cluster coordinates to find the nearest match with a tolerance of \rr1{$0.00015\deg$, $\sim2$} pixels of DECam. After that,
the CG angle \rr1{$\psi$} can be calculated using the ellipticity components (the \texttt{arctan2} function for $\arctan$): $\psi = \arctan(e_2/e_1)/2$; this angle is measured from the positive X-axis direction anticlockwise (top North, left East). \rr1{The CG angle is therefore defined as the angle between the major axis and the X-axis direction.}

Likewise, to test the robustness of the CG angle, we also measure it using the LS DR9 catalogues. We consider all galaxy model types and skip the photo-z cuts.
The signs of the ellipticity components need to be inverted to be consistent with the coordinate system above.
The two methods give compatible results (Section \ref{sec:robustness_CG}).

Next, we select galaxies in the shape and photometry catalogues that are $\pm30$ degrees around the directions (both side) of the major and minor axes of the CG using the position angle $\varphi$ to build two samples. We use this angle range to balance the number of selected galaxies for statistics and the signal variation as the angle changes. We test and find that using a smaller angle range, e.g. $15 \deg$, produces a larger difference between the stacked signals along the two directions but larger error bars as well. \rr1{On the other hand, a larger range (such as $45 \deg$) produces a smaller difference (as the triaxiality signature is diluted) but  smaller error bars. }

\begin{figure}
    \centering
    \includegraphics[width=\columnwidth]{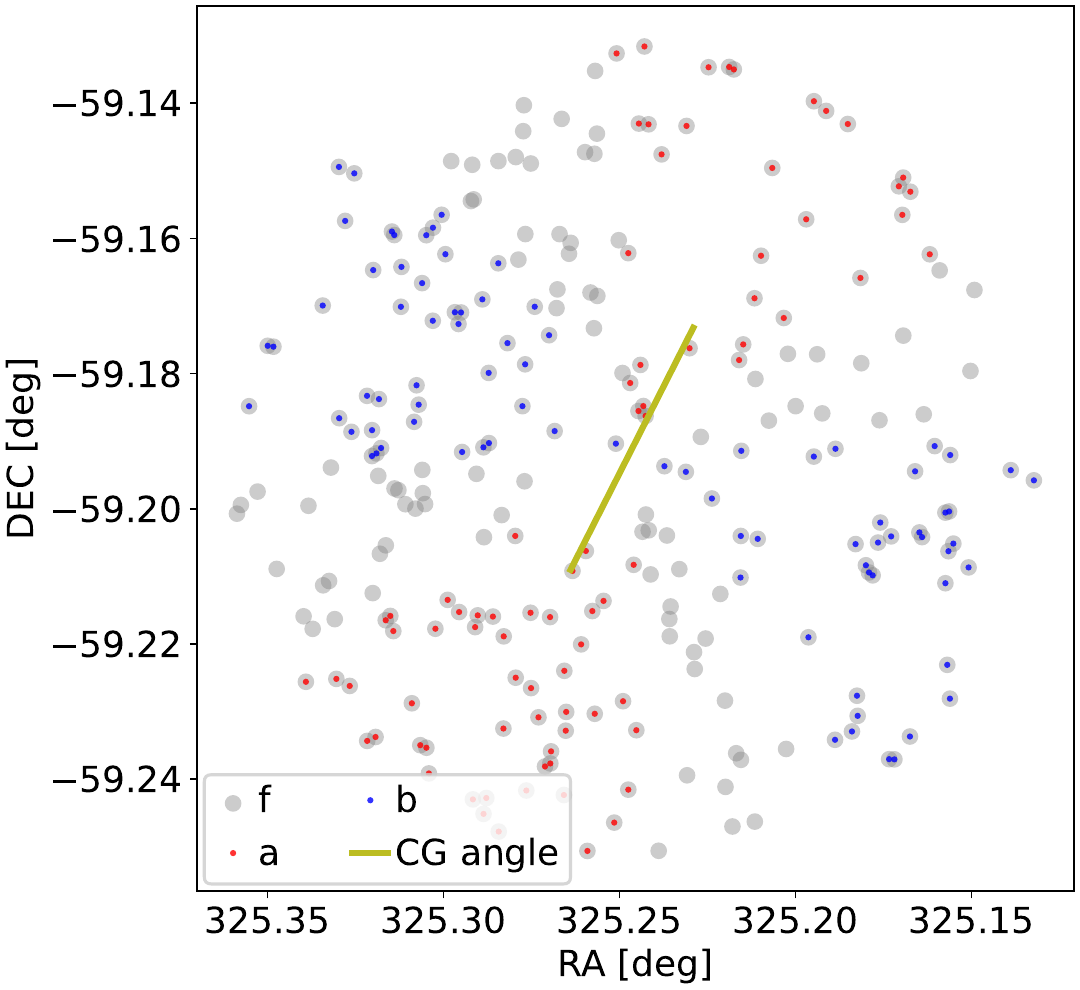}
    \caption{ 
    Example of making cuts for galaxies that are located around ($\pm 30\deg$) the directions of the major axis and the minor axis of the CG (DES Y1 \rdmp cluster ID 503; $z=0.34$) in the selected Y3 shape catalogue for lensing analysis. 
    For demonstration, we use a radial distance cut at $0.06\deg$ (1 Mpc).
    The red ($a$) and blue dots ($b$) show the major- and minor-axis samples respectively.
    The grey circles ($f$) show the source galaxies at all position angles.
    The yellow line is centred at the CG and shows the major axis direction of the CG. 
    Different from Figure \ref{fig:elliptical_halo_model}, here we use the CG's 2D orientation to select the samples since the halo orientation is unknown.
    We apply the same position angle cut to the photometry catalogue.
    }
    \label{fig:CG_angle_cut}
\end{figure}

Figure \ref{fig:CG_angle_cut} shows an example of how we make cuts based on position angles. 
Finally, we build profiles for \rr1{the galaxies selected by their position angles. The profile construction method is the same as the one for galaxies at all position angles (Section~\ref{sec:shape_catalog} and~\ref{sec:photometry_catalog}). }

\subsection{Cluster stacking and division}\label{sec:catalog_stacking}

We combine the signals from individual clusters to improve the statistics. 
Instead of combining the catalogues, we stack the cluster profiles to simplify the cluster selection (e.g. the binning of redshift and richness) and to reduce the computing resources; the two methods produces similar results. 

When we stack the lensing profiles, we use the jackknife resampling method~\citep[][and references therein]{Norberg2009,mcclintock19} to estimate the error bar of the stacked (mean) result. Each profile is weighted by the reciprocal of the scatter.

We use \texttt{kmeans\_radec}\footnote{\url{https://github.com/esheldon/kmeans_radec}} to divide the cluster sample into 100 similar regions, and each region roughly contains the same number of clusters. 
The \texttt{kmeans\_radec} method runs the K-means algorithm on the unit sphere and addresses the cosmic variation and LSS.

In jackknife \rr1{resampling}, each time we skip one region and compute a stacked profile. We compute the covariance of those stacked profiles (the values in radial bins) and multiply the covariance by the number of regions minus 1 \rr1{as the following}:
\begin{equation}
    \Bar{s}(R_i)=\frac{1}{K}\sum_{k=1}^K s_{\cancel{k}}(R_i)~;
    \label{eq:jackknife1}
\end{equation}
\begin{equation}
    C_{ij}=\frac{K-1}{K}\sum_{k=1}^K [s_{\cancel{k}}(R_i) - \Bar{s}(R_i)] [s_{\cancel{k}}(R_j) - \Bar{s}(R_j)]~. 
    \label{eq:jackknife2}
\end{equation}
\rr1{Here,} $s$ is a stacked profile, $K$ is the number of regions while $\cancel{k}$ means removing  the $k$th region. $i,j$ are the indices of radial bins\rr1{, and $C_{ij}$ is the covariance term between the $i$th and $j$th radial bins}.

We take the square root of each covariance matrix diagonal term as the stacked profile error bar at the corresponding radial bin, and we use the resampling mean ($\Bar{s}$) as the stacked profile value, which is very close to the direct mean of all profiles.
We use the same method to stack the number density and excess number density profiles without weight. \rr1{The} stacked profiles of all clusters \rr1{are presented} in Section~\ref{sec:fiducial_profiles}. 

We note that a few studies scale the radial distance by a cluster radius, e.g. the virial radius, before stacking. The advantage is that the clusters will be stacked at a `similar' size and thus the result can have better statistics, but the disadvantage is that this hinders comparisons with studies that use physical lengths and different cluster detection techniques, and also introduces a measurement uncertainty. We will test this method in future work.

In addition, we divide the clusters along the richness (a mass proxy) or redshift direction with two bins to investigate the richness and redshift dependence of the results; the stacking is performed at each bin (Section~\ref{sec:richness_redshift}). 

For the richness division, we split the clusters into two bins so that the sum (of the richness) in the two richness bins are comparable. This results in more low clusters in the low richness bin, but produces more comparable SNRs in the measurements of both bins.

For redshift, however, we use the median value of cluster redshifts to split them into redshift bins. In theory, there will be more massive clusters (high richness) at low redshift because of merging, but massive clusters at high redshift  are  easier to be detected. Thus we expect similar richness distribution in the low and high redshift subsamples. 

In Appendix \ref{sec:individual_bin}, we show an example of dividing the cluster sample into $2\times2$ subsamples by both redshift and richness. The results have larger statistical noise, and thus in the main text we consider 2 bins on redshift or richness only.

Finally, we use a flow chart to outline our pipeline and summarize our method in Appendix \ref{sec:flow_chart}.

\section{Results}\label{sec:results}

In this section, we present our results as follows.  
First, we examine optical measurements of our galaxy cluster sample through stacked weak lensing and galaxy (excess) number density profiles.  
To extract signatures of triaxiality, we examine \rr1{and compare \textit{axis-aligned profiles}, defined as the stacked} radial profiles \rr1{in azimuthal slices} along the \rr1{CG} major and minor axes, \rr1{and we normalize them by a \textit{fiducial profile}, which is a (randomly oriented) stacked profile that includes galaxies located at all position angles, to make the comparison more clear}.  
We also relate these measurements to an `effective splashback' radius, extracted from \rr1{the fiducial profile}.  
Then, we quantify any dependencies of triaxial signatures on the redshift or richness of galaxy clusters from our sample.  
Finally, we assess any differences in signature on the galaxy number density profiles that might depend on galaxy colour selection.

\subsection{Weak lensing and galaxy density radial profiles}\label{sec:fiducial_profiles}
\begin{figure*}
    \includegraphics[width=\columnwidth]{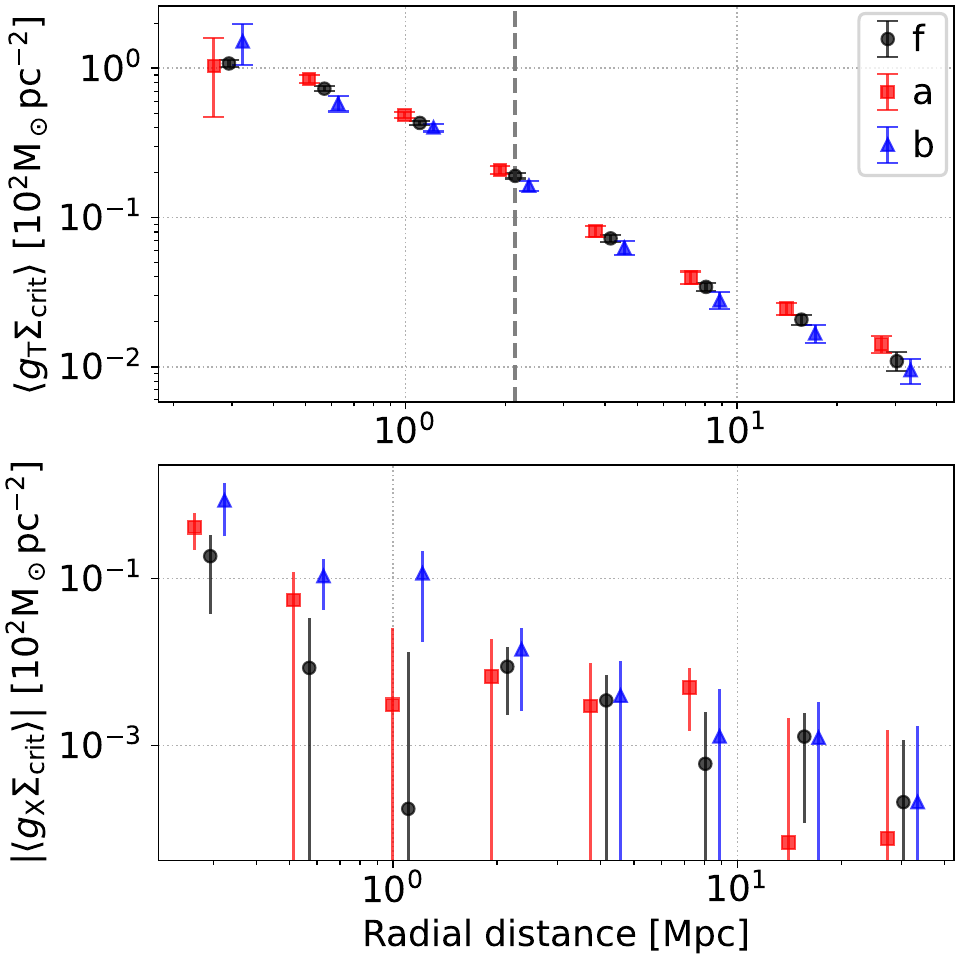}
    \includegraphics[width=\columnwidth]{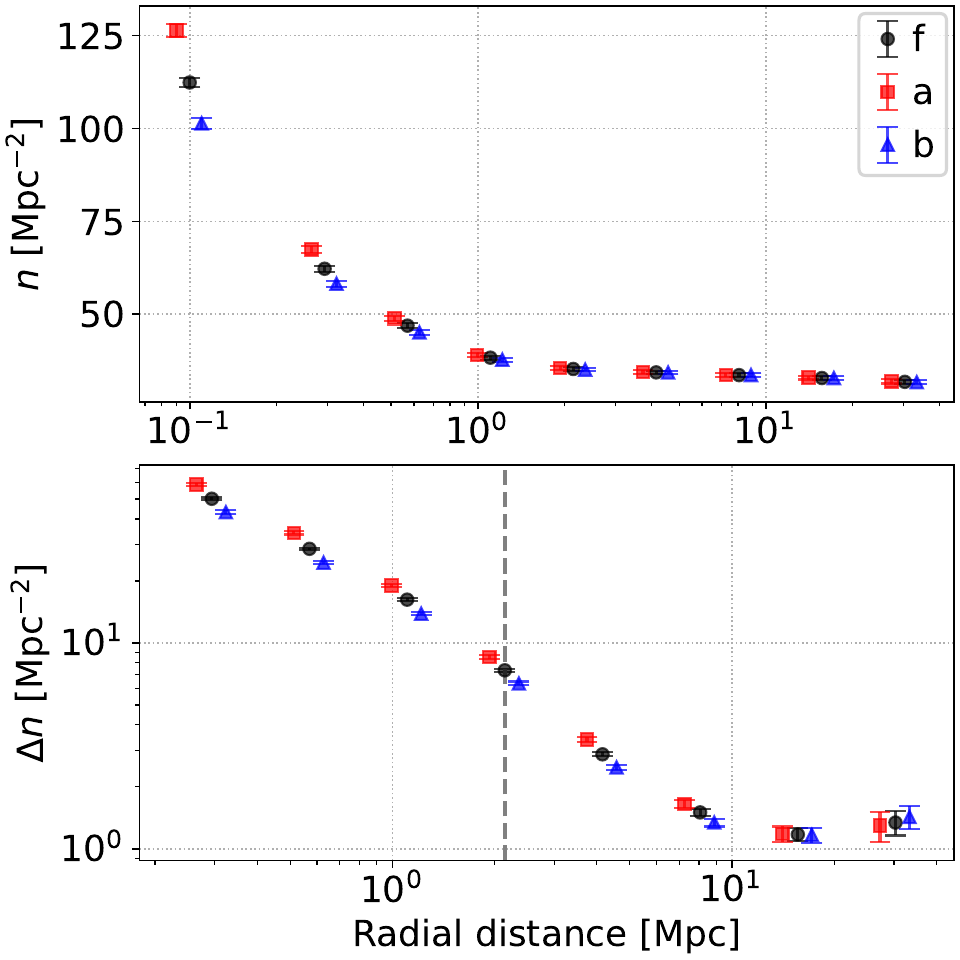}
    \caption{
    Stacked profiles from our main/canonical 
    dataset of DES Y1 \rdmp clusters ($\max\{\texttt{P\_CEN}\}\geq0.8$ for each) across all available redshifts ($0.2<z<0.86$)
    and 
    richness ($20<\lambda<235$):
    tangential excess surface mass density (\textit{top left}), cross excess surface mass density (\textit{bottom left}), galaxy number density (\textit{top right}), differential/excess galaxy number density (\textit{bottom right}). \rr1{The lensing mass density and galaxy number density are derived from DES Y3 and DES DR2 respectively.} 
    Here $f$ is the fiducial profile measurement using galaxies from the entire azimuthal range (4289 clusters; 11 out of 4300 are not in this footprint), while $a$ and $b$ show the samples around the directions of the major and minor axes of the central galaxy of individual clusters (4217 clusters; 72 out of 4289 failed to match the shape catalogue to obtain the central galaxy angle due to ambiguity e.g. blending). In all cases with sufficient signal-to-noise ratios for small enough error bars, measurements along the major axes exceed that of those made for the entire azimuthal range, and measurements along the minor axes are less than that made for the entire azimuthal range.   
    The vertical dashed line in each figure gives the local minimum of the logarithmic derivative of the fiducial profile of the cluster sample.
    For visualization, we add caps to the error bars in some graphs, and we shift the $a$ and $b$ points horizontally by 10\% but fix the $f$ points; the X-coordinates are bin centres  (same below).
    }
    \label{fig:canonical_dataset}
\end{figure*}

We first illustrate the stacked radial profiles of our main/canonical dataset, and differences in profile measurements due to triaxial shape. For most of our analysis, we use publicly available DES Y1 \rdmp clusters \citep{mcclintock19}. 
Figure~\ref{fig:canonical_dataset} shows the radial profiles of clusters subselected with $\max\{\texttt{P\_CEN}\}\geq0.8$.  This parameter indicates the confidence of the identified central galaxy (CG), which are more easily identified for more relaxed galaxy clusters.  We therefore use \texttt{P\_CEN} as a rough proxy for dynamical state; larger values of $\texttt{P\_CEN}$ also make it more likely that the orientation of the CG better aligns with that of the underlying dark matter halo.  As described in Section~\ref{sec:cluster_sample}, the entire sample has richness range across $20<\lambda<235$ and redshift range across $0.2<z<0.86$.  For all panels in Figure~\ref{fig:canonical_dataset}, the black circles correspond to profile measurements using galaxies from the entire azimuthal range.  The red squares and blue triangles respectively correspond to profile measurements in azimuthal slices (circular sectors) along the identified major and minor axes\rr1{, which we refer to as `axis-aligned' profiles}.  We use an angle cut of $\pm 30\deg$ on both sides for measurements in azimuthal slices.  Note, we shift data points by 10\% along the X-axis in each radial bin for visualization purposes.

The left column shows the (effective) excess surface density measurements, given by the lensing distortion in the tangential (top) and cross (bottom) directions with respect to the identified CGs as the galaxy cluster centres. 
The tangential measurement shows a relatively clean monotonic decrease with cluster-centric radius.  At $\sim2$~Mpc, we see a subtle change in the profile slope of the tangential signal.  This feature is analogous to the splashback feature observed in simulations~\citep{diemer2014,xhakaj2020} and real data~\citep{baxter2017,chang2018}.  For measurements in the latter, authors fit models to the full density profile instead of simply extracting a position of steepening from the excess surface density as is done here.

These profiles also exhibit a systematic relative strength of signals at each radial bin, aside from the innermost bin, which has the largest noise for the azimuthal slices.  The major axis ($a$) has the largest value, then the entire profile ($f$), then the minor axis ($b$).  The cross signal is $\sim10$ times smaller than the tangential and significantly noisier, showing no clear systematics, though exhibits a similar monotonically decreasing trend with radius; the relative strengths of the cross signal in each radial bin is much more difficult to disentangle due to the error bar size. 

The right column shows number density profiles. 
The top right profiles correspond to the \textit{projected} number density of galaxies $n(R)$.  Here, the error bars are small enough at all radial bins to see the systematic relative strength of signals even in the left-most radial bin.  Again, the measurement limited to the azimuthal slice along the major axis is the largest, then the measurement for the entire azimuthal range is in the middle, while the measurement for the slice along the minor axis is the smallest.  We see a precipitous drop in number density until $\sim1$~Mpc, which encompasses cluster member galaxies.  The bottom right profiles correspond to the \textit{excess number density} of galaxies, which we define as $\Delta n(R)=\overline{n}(<R)-n(R)$.  The excess number density is a closer analog to the lensing signal, as it reduces the contribution from foreground and background galaxies.  Here, we see a more continuous decline until $\sim10$~Mpc, and a subtle change in slope around $\sim2$~Mpc that plausibly corresponds to the same splashback feature in the tangential excess surface mass density profile.  The systematic relationship between the different azimuthal slices is consistent within radial bins inside the $\sim10$~Mpc range.  Outside of this, we expect stacked large scale structure traced by galaxies to have more azimuthal symmetry.  
\rr1{Later in Section~\ref{sec:red-sequence}, we examine these profile trends by subselecting red-sequence galaxies.}

\subsection{\rr1{Axis-aligned profile} trends}\label{sec:axes_ratios}

\begin{figure*}
    \includegraphics[width=\columnwidth]{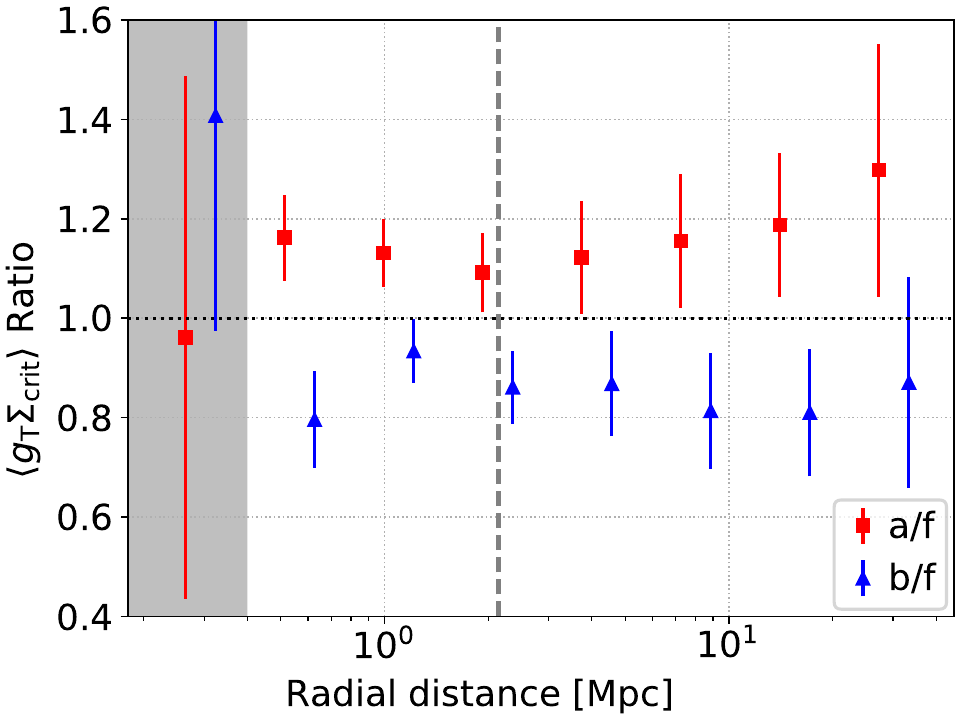}
    \includegraphics[width=\columnwidth]{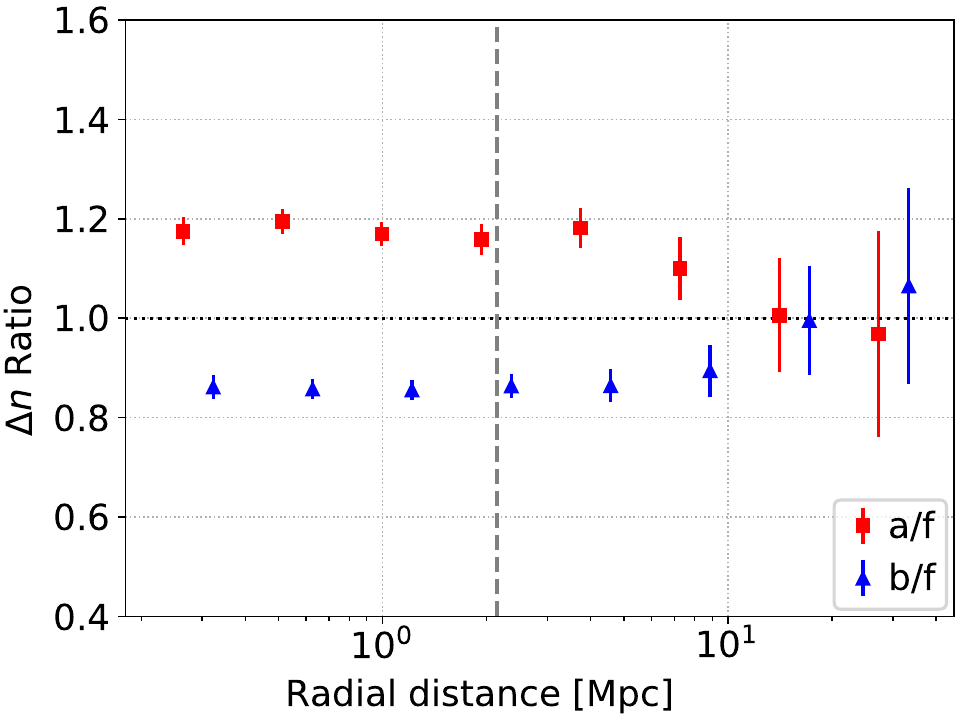}
    \caption{Normalized \rr1{axis-aligned} tangential excess surface density (\textit{left}) and the excess galaxy number density (\textit{right}) profiles for the canonical dataset (same data as shown in Figure~\ref{fig:canonical_dataset}).  We measure these along the major (red squares) and minor (blue triangles) axes of the central galaxy of each galaxy cluster and \textit{normalize} to the measurement made across the entire azimuthal range (fiducial). These correspond to ratios between the \rr1{(stacked)} profiles presented in Figure~\ref{fig:canonical_dataset}.
    In each panel the grey vertical dashed line shows the location of the local minimum of the logarithmic derivative of the respective fiducial profile (effective splashback radius).
    }
    \label{fig:canonical_dataset_ratios}
\end{figure*}

We now discuss relative features in \rr1{axis-aligned} profiles, measured in azimuthal slices aligned with the major and minor axes. \rr1{The slices span $\pm30$ degrees, centred on the CG major ($a$) or minor axis ($b$).}   Figure~\ref{fig:canonical_dataset_ratios} illustrates the profiles measured along the major axis (red squares) and along the minor axis (blue triangles), each normalized \rr1{with respect to the fiducial profile ($f$) measured for the entire azimuthal range (the measurement made with galaxies that are at all position angles)}, with error bars estimated by uncertainty propagation. 

The left panel of Figure~\ref{fig:canonical_dataset_ratios} shows the normalized excess surface density profiles.  These profiles probe the projected mass distribution.  Aside from the innermost radial bin, which has very large error bars due to the low lensing SNR produced by limited source galaxies, the major axis measurements, $a$, are larger than all measurements made for the entire azimuthal range, $f$.  Data points of $a/f$, including errors, all sit above a ratio of 1.  The reverse is true for the measurements along the minor axis.  All $b/f$ ratios including the error bars (except for the measurement in the innermost radial bin), mostly sit below a ratio of 1. The offsets from 1 in $a/f$ and $b/f$ are both $\sim10-20\%$.  

Interestingly, we see hints of a `necking' feature; the profiles exhibit a radial trend where $a/f$ and $b/f$ data points converge towards 1 at a radial distance of about $\sim1-2$~Mpc, before separating again.  We see hints of this effect with other datasets and also in subsamples of galaxy clusters in narrower richness and redshift bins (see Section~\ref{sec:richness_redshift}).  
The location of this `neck' is consistent with simulation measurements of surface mass density/lensing signal of triaxial haloes made by \citet{osato2018} and \citet{zzhang2023}, who also found a `neck' feature around the transition between the one-halo and two-halo dominated regimes. However, their methods to extract triaxial signatures are different from ours -- they considered source galaxies at all position angles when fixing the angle between the halo major axis and the line of sight,  
while we consider signals along the CG (projected) major- and minor-axis directions only.  
Our measurements may be the first observational evidence of this `necking' feature existing in the triaxial mass distribution around clusters.
In the left panel of Figure~\ref{fig:canonical_dataset_ratios}, the similarity in \rr1{both axes-aligned} measurements at $\sim1$~Mpc is likely due to this location corresponding to an effective edge of the clusters associated with the approximate average splashback locations.  Outside of $\sim1$~Mpc, cluster-feeding filaments govern the mass distribution.  Filaments likely align with the \rr1{orientation} of the triaxial cluster (and the CG), leading to another boost in the separation between the two \rr1{axes-aligned profiles}~\citep{tempel2015, Gouin2020}.

The right panel of Figure~\ref{fig:canonical_dataset_ratios} shows the normalized excess galaxy number density profiles.  While luminous galaxies trace underlying mass distributions, they are biased tracers and we do not expect a perfect correspondence \citep{kauffmann1997, coil2013}.  The error bars of these normalized profiles are much smaller than those in the normalized tangential excess surface density profiles.  Here, we see an almost flat offset until about $\sim8$~Mpc.  Inside this radius, the profile measured along the major axis is $\sim20\%$ larger than that of the profile measured using all galaxies and the profile measured along the minor-axis is $\sim15\%$  smaller than that of the profile measured using all galaxies.  

Compared with the \rr1{normalized axis-aligned} profiles of the tangential excess surface density, the values \rr1{normalized in the excess galaxy density measurements }are similar. But, the `necking' feature in the \rr1{normalized axis-aligned }excess galaxy density profiles occurs at a much larger radius of $\sim10-20$~Mpc, well outside the environments of each galaxy cluster. Ratios \rr1{at larger radii} converge to unity quickly. 
Overall, the `necking' feature corresponds to the average radial location where \rr1{axis-aligned profile }measurements appear to become more similar to one another.  
The `neck' location occurs near the effective splashback location for the tangential excess surface density measurements (between $\sim1-5$~Mpc when we vary the richness), while the feature for the excess number density measurements does not appear to have the same association.  We see hints that the ratio offset slightly decreases near the effective splashback radius derived from the fiducial excess number density profile.
We further discuss these in Section~\ref{sec:richness_redshift}, where we examine how redshift and richness impact these features. 

We note that these observational trends reasonably correspond to those described in the toy model of an elliptical NFW profile with axis ratio 2/3, laid out in Section~\ref{sec:theory}, although an elliptical two-halo term may be necessary for explaining the `necking' feature. The result illustrated in Figure~\ref{fig:elliptical_halo_model} also shows a relatively flat offset in \rr1{ratio values} by $\sim15\%$ around 1 in tangential reduced shear and excess convergence. 

\subsection{`Effective splashback': Transition locations in projected profiles}\label{sec:splashback}

Here, we illustrate features in the \rr1{stacked} profiles that correspond to a {\it transition in the logarithmic  derivative} of each profile of the optical signature.  We call this an `effective splashback' (in 2D), as this is a similar method used by \cite{more2016,chang2018,sunayamamore2019}.  
However, previous works have identified the splashback radius from {\it deprojected} profiles to model the 3-dimensional mass or galaxy density distribution.  
We illustrated the transition location in the projected surface mass density profiles and the excess/differential galaxy number density profile, respectively as vertical dashed lines.  Note, the location of the `effective splashback' location occurs in the same radial bin for both measured profiles.  Given the coarseness in radial binning, this is consistent with the results from~\cite{chang2018}, who found agreement between the 3D splashback locations derived from galaxy density and lensing data respectively; 
note they used comoving coordinates to improve the detection of the 3D splashback feature\footnote{The 3D splashback radius scales with $r_\textrm{200m}$~\citep{diemer2014}, which scales with $1/(1+z)$ when $M_\textrm{200m}$ is fixed.}, but we use physical distances since the goal of this paper is to study cluster triaxiality. \rr1{Also, our sample includes richer (thus more massive) clusters, which have larger splashback radii.}

In Figure~\ref{fig:canonical_dataset_ratios}, we can compare the effective splashback locations with trends in the normalized \rr1{axis-aligned} profiles -- $a/f$ and $b/f$ ratios.  In the left panel, the \rr1{ratio values} for the excess surface density exhibit a `necking' feature near the effective splashback radius, at a radius of $\sim1-2$~Mpc.
In the right panel, a similar feature for the profile ratios of $\Delta n$ occurs at around $\sim10-20$~Mpc.  
\rr1{
Plausible reasons for the different `neck' locations in the two normalized axis-aligned profiles are as follows. We can attribute the ‘necking’ feature in lensing profiles to projection effects from nearby low-mass regions associated with randomly distributed LSS. On the other hand, we can attribute the `necking' feature in number density profiles to the low-mass regions at very large radii. Galaxies that are detectable (thus sufficiently bright and massive) are less likely to reside in those low-mass regions. 
}
The distribution of the \rr1{accreting} galaxies likely \rr1{sits} in \rr1{primary} filaments feeding into the cluster\rr1{; those primary filaments align with the cluster halo and are more massive than the neighboring randomly distributed LSS}.  
\rr1{Thus}, both the lensing `necking' feature and the effective splashback locations more closely indicate the edges of the cluster mass distribution.
\rr1{We give a more detailed explanation in Section~\ref{sec:richness_redshift}.}

\begin{figure*}
    \includegraphics[width=\columnwidth]{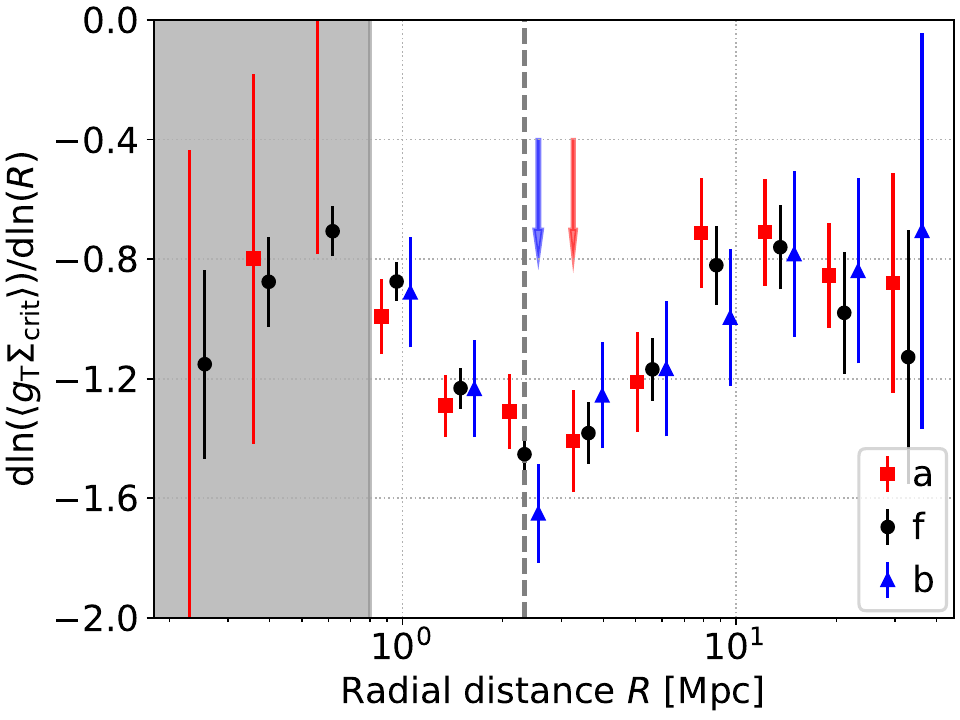}
    \includegraphics[width=\columnwidth]{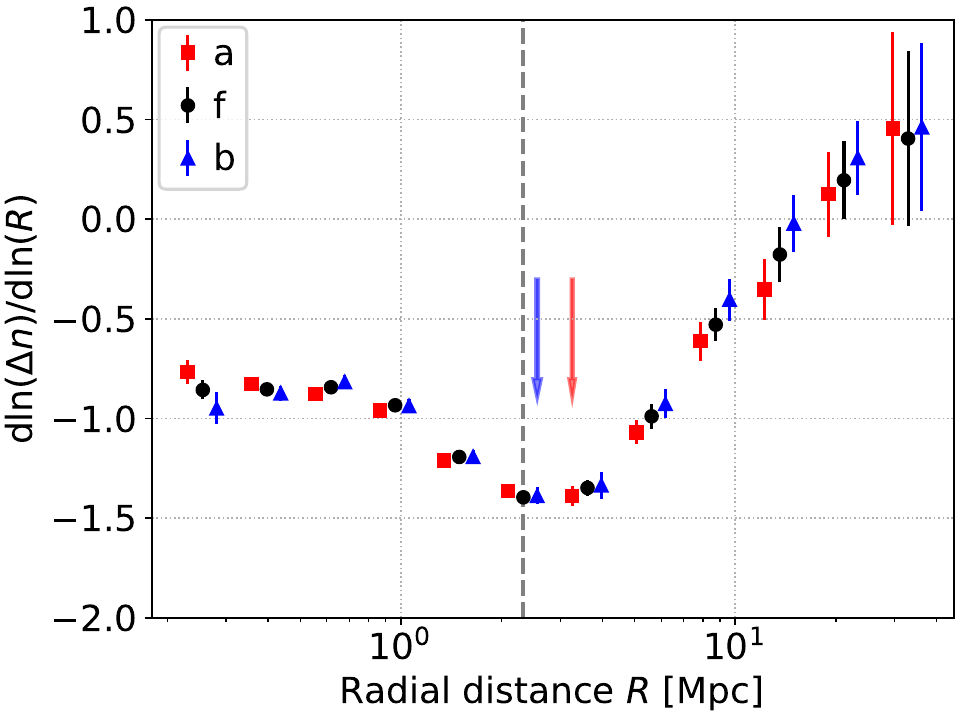}
    \caption{\textit{Left}: Logarithmic derivative profile of the effective excess surface density. Because $\langle g_\textrm{T} \Sigma_\textrm{crit} \rangle$  of the minor-axis sample ($b$) is negative at the second radial bin, we omit the values at that bin and neighbouring bins.    \textit{Right}: Logarithmic derivative profile of the excess number density.   
    The vertical dashed line in each panel indicates the local minimum for the fiducial sample between 1 and 10 Mpc, corresponding to an `effective splashback' location. 
    The red and blue arrows indicate the local minimums for the major-axis sample and the minor-axis sample respectively.
    }\label{fig:effective_splashback_calculation}
\end{figure*}

To identify the `effective splashback' in each profile, we calculate the derivative over 12 radial bins illustrated in Figure~\ref{fig:effective_splashback_calculation}.  Note, we calculate the effective splashback with 12 instead of the 8 radial bins shown in the figures from Section~\ref{sec:fiducial_profiles}.  While the increase in the number of radial bins (by 50\%) worsens the noise, particularly for the excess surface mass density profiles, this allows us to identify a more accurate location of a `dip' in the profile \rr1{derivative} ($\sim2$~Mpc). 
There are signs that the local minimum of the minor-axis sample ($\sim2$ Mpc) is smaller than the one of the major-axis sample  ($\sim4$ Mpc) both in the lensing and excess number density ratios (as indicated by the arrows in Figure~\ref{fig:effective_splashback_calculation}), which physically makes sense since a cluster is expected to be `longer' along its major axis; but these phenomena could be caused by noise as well.

\subsection{Triaxial signature dependency on redshift and richness}\label{sec:richness_redshift}
\begin{figure*}
    \includegraphics[width=1.95\columnwidth]{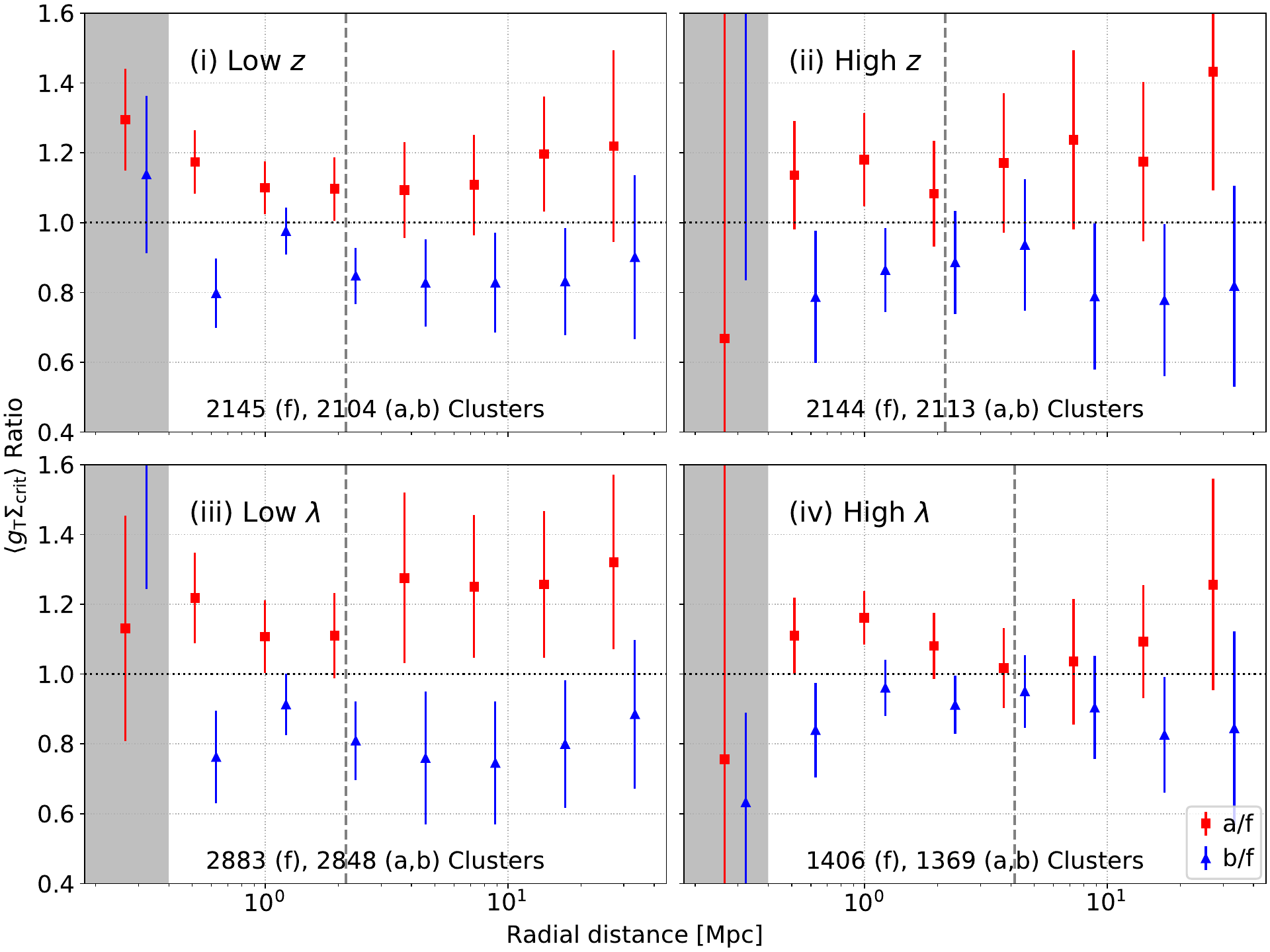}
    \caption{
    \rr1{Normalized axis-aligned profiles showing the }ratios of the excess surface mass density for subsamples of our canonical dataset.  \rr1{Major axis profiles are in red squares, and minor axes profiles are in blue triangles.}
    We plot both results for the tangential lensing distortion and the excess number density (next figure) for comparison.
    The \textit{top left (i)/right (ii)} panels show the difference between low/high redshifts, while the \textit{bottom left (iii)/right (iv)} panels give the difference between low/high richness.
    The `necking' feature of both left-hand panels (low redshift and low richness) occur at smaller radii compared with the locations of that feature in the right-hand panels.  Annotations indicate the number of clusters used in the redshift or richness bin for the `effective splashback (dashed line)' measurement ($f$; fiducial), or for the ratio profile measurement ($a, b$; major, minor axes).  Not all clusters have a clear CG to use for axis identification due to blending and/or the presence of multiple CGs.
    }
    \label{fig:bins_lensing}
\end{figure*}

In this subsection, we examine how the \rr1{normalized axis-aligned profiles} discussed in Section~\ref{sec:axes_ratios} behave in different richness and redshift ranges.  Figure~\ref{fig:bins_lensing} shows normalized profiles of the lensing distortion, defined by the tangential excess surface density.  
We subdivide our cluster sample from Figure~\ref{fig:canonical_dataset} into only two subsamples in order to preserve sufficient SNR for a measurable difference between the major- and minor-axis measurements\rr1{, respectively red squares, indicated as $a/f$ and blue triangles, indicated as $b/f$}. 

The top row illustrates the difference between a lower redshift subsample (top left: $0.2<z<0.47$) and a higher redshift subsample (top right: $0.47<z<0.86$).  The radial location of the `neck' in the low redshift subsample is consistent with that of the entire sample, shown in Figure~\ref{fig:canonical_dataset_ratios} at $R_\mathrm{neck}\sim1-2$~Mpc.  This is not surprising, given that the low redshift subsample likely drives most of the signal in the total stack.  
The high redshift subsample unfortunately has too large error bars to clearly show the location of the `neck', but may be anywhere between the third and fifth radial bin (excluding the shaded bin; $R_\mathrm{neck}\sim2-8$~Mpc).   

In contrast, there is a clearer difference in the location of the `neck' between low and high richness objects.  The bottom row illustrates the difference between the lower richness subsample (bottom left: $20<\lambda<33$) and the higher richness subsample (bottom right: $33<\lambda<235$).   The neck in the low richness subsample is consistent with that of the entire sample at $R_\mathrm{neck}\sim1-2$~Mpc.  However, the feature in the high richness subsample occurs closer to  $R_\mathrm{neck}\sim4-5$~Mpc. We have also included the corresponding `effective splashback' locations derived by respective fiducial profiles for each subsample shown in the panels.  We note that the effective splashback location for the high richness subsample is also at a larger radius ($\sim4$~Mpc), similar to the `neck'.  The larger radial locations of both the neck and the effective splashback feature are consistent with the fact that the larger richness subsample has more massive clusters with larger characteristic radii.

One additional reason for the location of the `necking' feature in the excess surface mass density ratios and the correspondence to the effective splashback location may be that the excess surface mass density is sensitive to projection effects from nearby correlated LSS along the line-of-sight.  These structures would be influenced by the galaxy cluster environment and therefore correlated with the galaxy cluster location.  But, these structures may not correspond to the primary filaments aligned with the cluster axes, and their orientations are more randomly distributed.  The random orientations introduce a rounding effect on the weak lensing profile ratios that produces the `necking' feature washing out the triaxial signature, but not the galaxy distribution.    
Beyond the radial location of the `neck', this LSS contribution to the mass along the line of sight becomes less significant, leading to an overall projected mass that aligns with the central halo; thus, the triaxial signature in the lensing signal recovers at larger radii. 
This `necking'/rounding feature in the normalized lensing profiles cannot be produced by the cluster halo and primary filaments only, because otherwise the normalized galaxy density should show a clear `neck' at the same location. It is not generated by uncorrelated LSS either, which is randomly distributed and would wash out the lensing triaxial signature quickly.

\begin{figure*}
    \includegraphics[width=1.95\columnwidth]{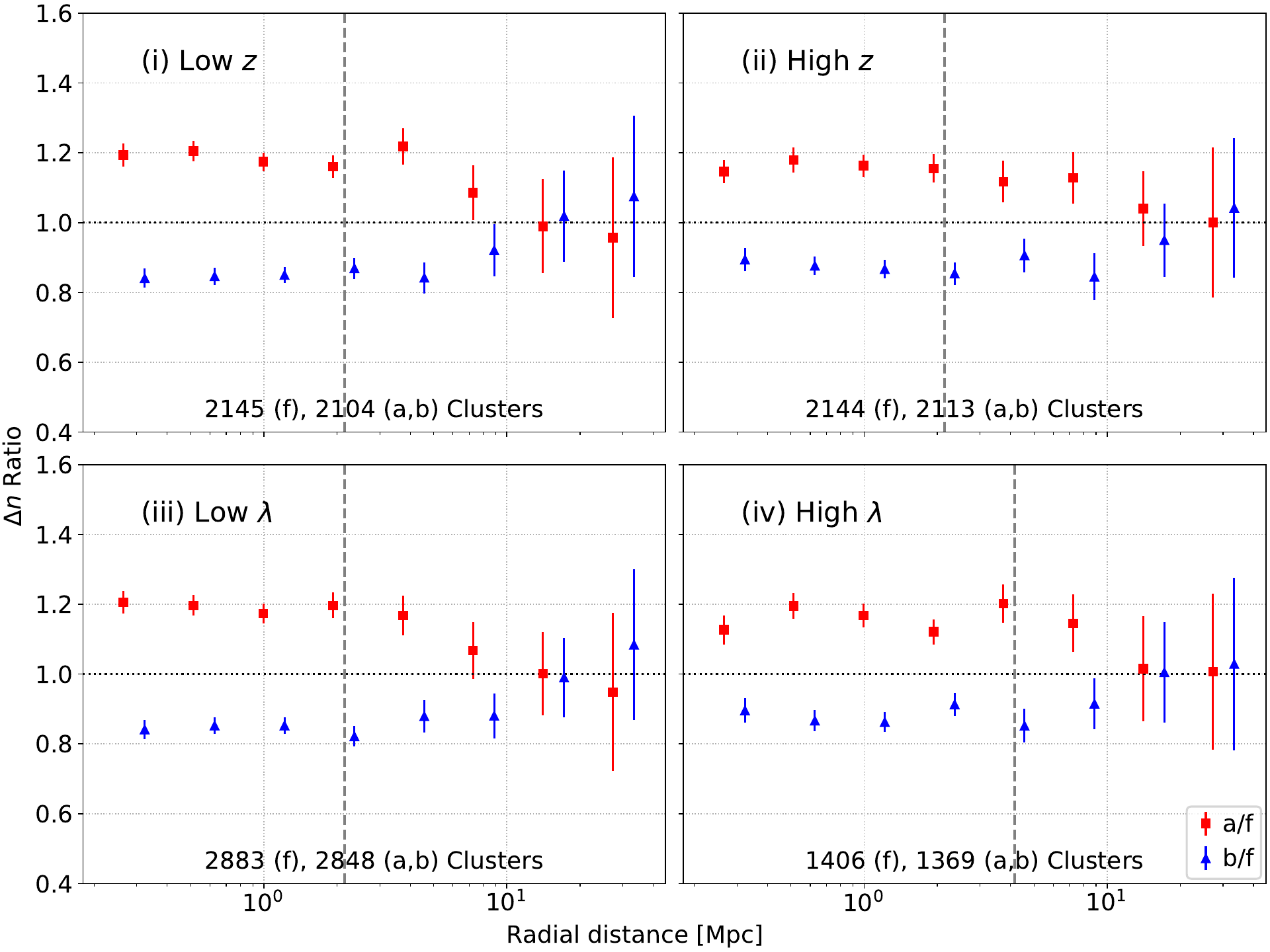}   
    \caption{
    \rr1{Normalized axis-aligned profiles showing the }ratios of the excess number density \rr1{for subsamples of our canonical dataset.} Each panel shows the measurement for the same subsamples as in Figure~\ref{fig:bins_lensing}: the \textit{top left (i)/right (ii)} panels show the difference between low/high redshifts, while the \textit{bottom left (iii)/right (iv)} panels give the difference between low/high richness.
    Again, low redshift or small richness \rr1{subsamples exhibit smaller} radii of \rr1{convergence}/`neck'. 
    }
    \label{fig:bins_number}
\end{figure*}

Figure~\ref{fig:bins_number} shows normalized profiles of the excess number density, measured in azimuthal slices along the major (red squares, indicated as $a/f$) and minor (blue triangles, indicated as $b/f$) axes.  Again, we normalize profiles with respect to the measurement made with all galaxies ($f$) in that radial bin.  The excess number density profiles have much smaller error bars compared with the excess surface density profiles.  The top and bottom rows correspond to the same respective redshift and richness subsamples as in Figure~\ref{fig:bins_lensing}. In Appendix \ref{sec:individual_bin} we consider a further division that splits clusters by both redshift and richness but the results are noisier. 
Here we also present the `effecitve splashback' location of the fiducial excess number density profile in each cluster subsample, which is still consistent with the lensing results.

While the `neck' locations in the tangential excess surface density measurements occur near the effective splashback location in each subsample, the similar `necking' feature occurs at {\it much} larger radii for the analogous measurement in excess number density.  The `necking' feature of the low redshift subsample occurs between the antepenultimate and the penultimate radial bins, $R_\mathrm{neck}\sim10$~Mpc. The low richness subsample has hints of $R_\mathrm{neck}\sim10$~Mpc as well.  
The locations of the `neck' in other subsamples are more closer to the corresponding location for the entire sample: they sit closer to the penultimate bin, with $R_\mathrm{neck}\sim10-20$~Mpc.  

These trends indicate that the excess number density measurements along each axis do not converge to one another until well outside the cluster region, at a radial distance of $\gtrsim10~$Mpc, regardless of effective splashback location.  In other words, our selection of cluster galaxies more strongly traces signatures of triaxiality to larger radii.  This is likely due to the fact that our galaxy selection for the excess number density profiles is comprised of galaxies associated with the galaxy cluster, both cluster members and accreting galaxies at the same redshift.  These galaxies would trace the cluster-feeding filaments that drive the underlying shape of the cluster.  

Note, the potential rounding effect from LSS that could contribute to the `necking' feature in the excess surface mass density ratio profiles would not affect the galaxy density.  Correlated structures not associated with the primary filaments feeding the galaxy cluster would be less likely to contain galaxies (at least bright ones) because of their low masses.  Thus, the `necking' feature in the galaxy distribution is solely determined by the primary filamentary structures (which are more massive) at even larger radii.    

Also, signatures of triaxiality seem to completely disappear after the `necking' feature in the galaxy distribution, whereas there is still a systematic offset in ratios at large radii for the excess surface mass density profiles. For the radial distances $\gtrsim10$~Mpc, galaxies that trace the DM filaments are likely lower mass objects below our detection threshold.  Thus, the triaxial signature of structure surrounding galaxy clusters disappears at the largest radii in the excess galaxy distribution but not in the excess surface mass distribution.  
In the future, we will re-examine these in cosmological simulations.

Despite the wide subselection in redshift and richness, we can summarize some (non)-trends for the lensing and excess number density results. 
First,  the `effective splashback' location does not have a clear dependence on redshift.  
Both subsamples divided by redshift have similar distributions in richness, and therefore similar distributions in mass  \citep{melchior2017,mcclintock19}.  Any trends would therefore be due to redshift dependence in non-mass dependent features of the subsamples, such as dynamical state.  While we expect higher redshift objects to be more dynamically unrelaxed, we do not see any measurable difference of `effective splashback' locations within the radial bins and redshift subsamples measured. We do find that the `neck' location seems to show up in a larger radius in the higher redshift subsample.  
Second, both the `effective splashback' location and the `neck' location occur at noticeably larger radii for the high richness subsample, consistent with the fact that high richness clusters are more massive with larger sizes and splashback radii.
The corresponding increase in the `neck' radii of the lensing signal is consistent with the assertion that the `neck' occurs near the average edge of the cluster mass distribution, where the tangential excess surface density along the major axis is more similar to that along the minor axis. For galaxy number density, this means a larger cluster tends to have longer filamentary structure connected to it.
Third, we do not see the same association of the physical scale between the `effective splashback' and `neck' location for the excess surface number density of galaxies as the excess surface mass density, likely due to the fact that our galaxy selection more closely traces the primary filaments and associated triaxial alignment near the cluster edge,  while the lensing signal is more affected by the projection effect.  
Finally, the ratio values (as a proxy of the mean halo triaxiality) do not strongly depend on redshift and richness. The high richness clusters seem to have ratios closer to 1 in the lensing results (i.e. rounder shapes) but the `necking' feature also affects the ratio, and the excess number density ratio does not show a similar change. Still the ratios along the two axes are nearly symmetric with respect to unity; this asymmetry is mainly due to the fact that the azimuthally averaged signal ($f$) includes regions in addition to the ones along the major ($a$) and minor ($b$) axes (Figure~\ref{fig:CG_angle_cut}).

\subsection{Red-sequence galaxies distribution}
\label{sec:red-sequence}

In this subsection, we examine how galaxy populations impact the signature of triaxiality on the \rr1{axis-aligned} excess number density profile measurements.  We split galaxies associated with each cluster into a red-sequence and non-red-sequence population, identified by the \rdmp colours and magnitude limits: for each cluster, we compute the median and the scatter of colour $r-i$ and $i-z$, and the magnitude limit in $r,i,z$ of the cluster member galaxies reported by \rdmp ($g-r$ has a larger scatter), and then select the galaxies in the DES DR2 catalogue (pre-selected by the method in Section \ref{sec:photometry_catalog}) using those magnitude limits and 1.5 times the standard deviation around the median of each colour. The reason for selecting galaxies in the DES DR2 catalogue rather than directly  using the \rdmp cluster member catalogue is that the member galaxies in the \rdmp catalogue have  a radial distance limit $R_\lambda\sim1$~Mpc, which is close to the virial radius~\citep[][and references therein]{mcclintock19}.     
Note, galaxies beyond the virial radius of each galaxy cluster are no longer cluster member galaxies, but are otherwise associated with the nearby filaments, and are likely to accrete.  

\begin{figure*}
    \centering
    \includegraphics[width=1.95\columnwidth]{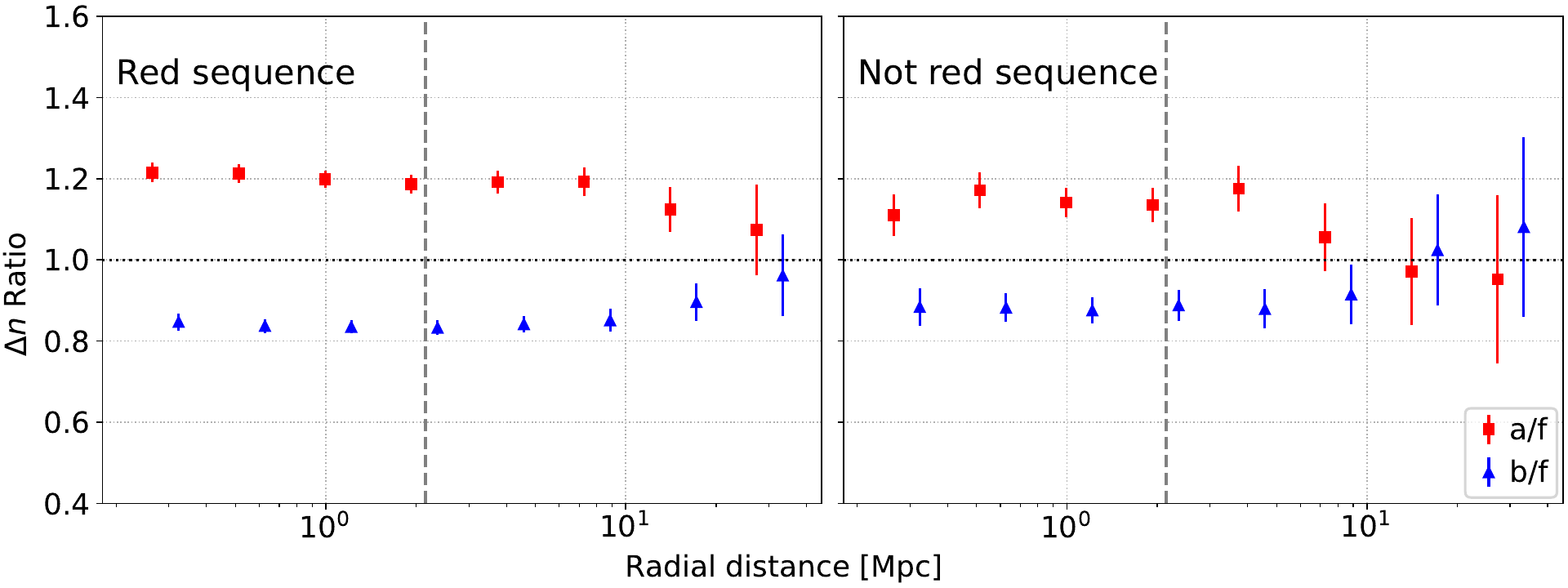}
    \caption{\rr1{Normalized axis-aligned profiles showing} ratios of excess number density for red-sequence (\textit{left}) and non-red-sequence (\textit{right}) galaxies. Note, the location of the `effective splashback' radial bin (the grey vertical dashed line) does not measurably change with the galaxy sample selection, but red-sequence galaxies retain triaxial signatures \rr1{in these profiles} to larger radii.}
    \label{fig:RS_ratio}
\end{figure*}

Figure~\ref{fig:RS_ratio} shows the excess number density ratios of the signal along the major (red squares) and minor (blue triangles) axes to the fiducial signal (generated by galaxies in all directions) as functions of distance from cluster centres.  The left panel shows this ratio for excess number density profiles measured with red-sequence galaxies associated with the galaxy clusters, and the right panel for profiles measured with all other associated galaxies.  The primary difference between the two is in the `neck' location.  For red-sequence galaxies, the \rr1{necking} feature occurs in the largest radial bin, $\sim30$~Mpc.  The \rr1{necking} feature in profile ratios from all other associated galaxies occurs closer to $\sim10$~Mpc.  
These results are consistent with measurements from \cite{zhang2013}, who found that red galaxies better trace large scale structure and intercluster filaments.  Specifically, luminous red galaxies (LRGs; or red-sequence galaxies in clusters) are commonly used tracers for  LSS and the matter distribution in our universe.  In cluster environments, the distribution of these objects trace triaxial signatures to larger radii than other galaxies.  Figure~\ref{fig:RS_ratio} illustrates that red-sequence galaxies in our sample do in fact trace the extended mass distribution of triaxial galaxy clusters to larger radii.

We note that the effective splashback location appears in the same radial bin for both profiles (and the same bin for the whole sample, Figure \ref{fig:canonical_dataset_ratios}), shown with the vertical dashed line.  
Interestingly, the shift in `neck' location does not track any shifts in the effective splashback location, contrary to the relationship between the two features in Figures~\ref{fig:bins_lensing} and \ref{fig:bins_number} for high richness clusters.    
The stability of the effective splashback radius is consistent with conclusion from \cite{baxter2017}, who showed that the location in profile steepening is consistent between the red and blue galaxies in their galaxy cluster sample considered. 
The effective splashback location of galaxies off the red sequence could come from a population of `green' galaxies that are quenching star formation as they enter the cluster and start to rotate around the cluster centre~\citep{shin2019}.

Additionally, we find the ratio values do not have a strong dependence on galaxy colours and are generally flat at small radii. The galaxies off the red sequence show a smaller offset from unity but with larger error bars. Again, the ratios along the two axes are nearly symmetric with respect to 1.

\section{Discussion}
\label{sec:discussion}

\subsection{Robustness of results}\label{sec:robustness}
Here we test the robustness of our results by adjusting the sources of catalogues (generated by different measurement methods) one by one. 
The same physical quantity should have similar results no matter what \rr1{measurement} methods are used.

\subsubsection{Robustness to lensing signature and number density measurements: Legacy Surveys shapes and photometry}\label{sec:robustness_LS}

\begin{figure*}
    \includegraphics[width=\columnwidth]{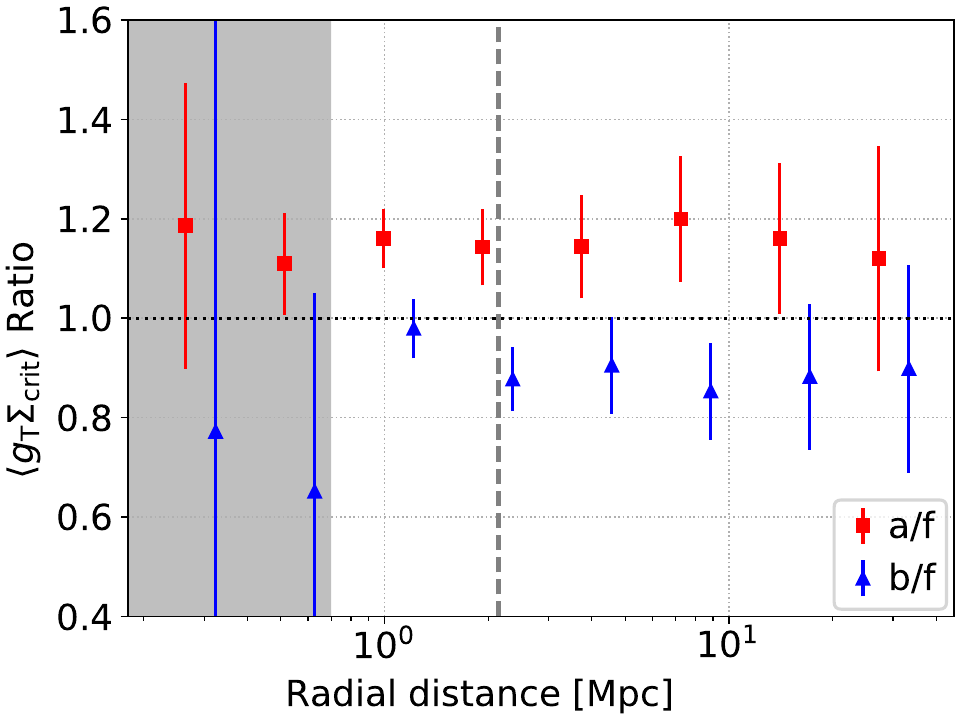}
    \includegraphics[width=\columnwidth]{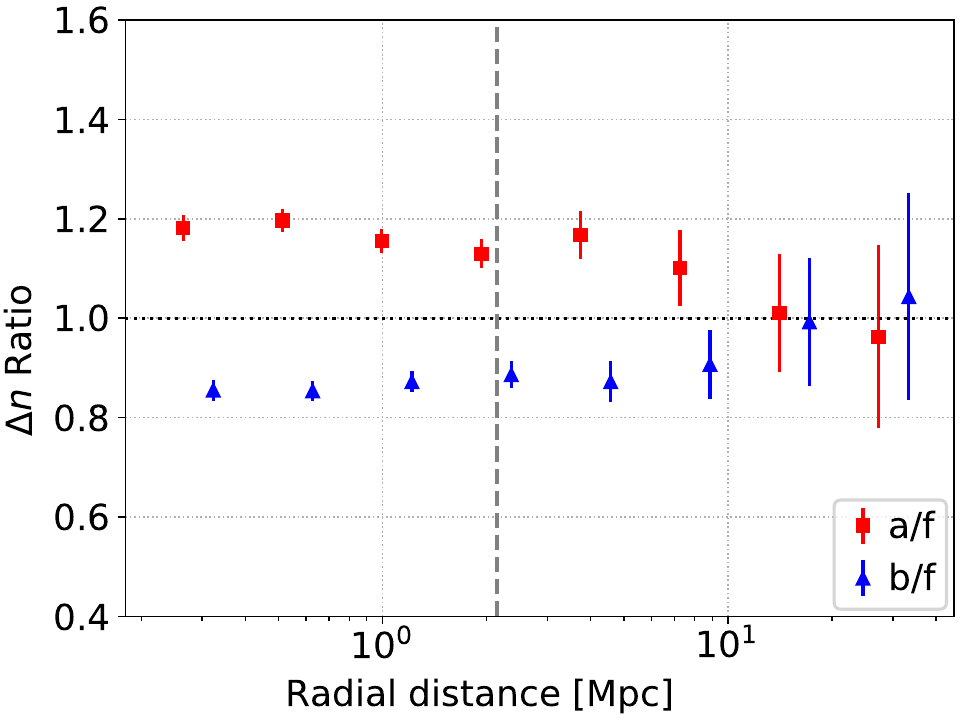}
    \caption{\rr1{Normalized axis-aligned profiles showing} excess surface mass density ratios (\textit{left}) and excess galaxy number density ratios (\textit{right}) using the Legacy Surveys DR9 catalogues but DES Y1 \rdmp cluster catalogue. The fiducial ($f$) profile measurement uses galaxies at all position angles (4300 clusters); the major- ($a$) or minor-axis ($b$) profile measurement uses 4217 clusters. The cluster sample is selected by $\max\{\texttt{P\_CEN}\}\geq0.8$ and spans richness $20<\lambda<235$ and redshift $0.2<z<0.86$.}
    \label{fig:LS_ratios}
\end{figure*}

We use the Legacy Surveys DR9\footnote{\url{https://www.legacysurvey.org/dr9/}} shape and photo-z measurements for building lensing profiles and Legacy Surveys DR9 photometry for building excess number density profiles, but still use the DES Y1 \rdmp cluster catalogue. The CG angles still come from the DES Y3 shapes.
In Figure \ref{fig:LS_ratios}, we show the \rr1{normalized axis-aligned profiles}, with the dashed vertical line showing the local minimum of the logarithmic derivative of each LS \rr1{fiducial} profile (both effective splashback locations are at $\sim2$~Mpc). 
The results are consistent with the ones derived from the DES Y3 shapes and DR2 photometry: a clear (and nearly symmetric) difference between the major- and minor-axis samples, with ratio values spanning $\sim10-20\%$ around 1; a probable `necking' feature at $\sim1-2$~Mpc in the excess surface mass density ratio (though the error bar of the minor-axis sample is large at the second radial bin); an excess number density ratio `neck' at $\sim10-20$~Mpc. 

We also test and find that using DES Y1 shape catalogue~\citep{zuntz18}, photo-z catalogue~\citep{hoyle18}, and DES DR1 photometry~\citep{abbott18} produces consistent but noisier results, and the reason is that DES Y3 \rr1{shape catalog} has improved measurement methods and \rr1{DES DR2 photometry catalog} has improved depth.

\subsubsection{Robustness across CG angle measurements methods}\label{sec:robustness_CG}

\begin{figure}
    \includegraphics[width=\columnwidth]{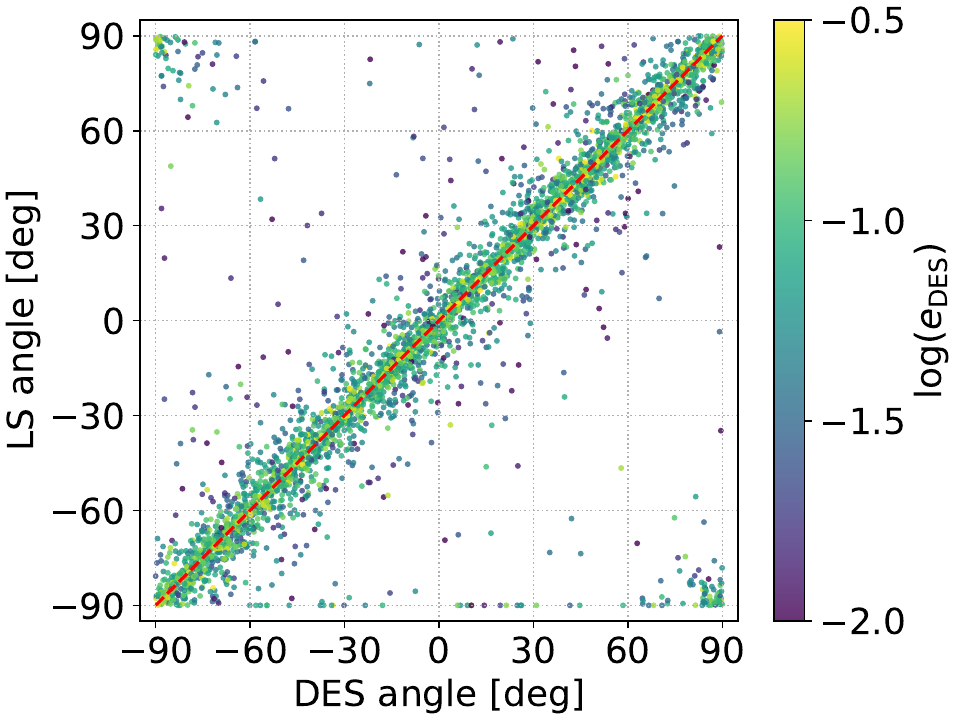}
    \caption{Comparison between the CG angles derived from the DES Y3 \metacal shapes (X-axis) and the Legacy Surveys (LS) DR9 Tractor model fit shapes (Y-axis), colour-coded by the logarithm to the base 10 of the DES ellipticity, to illustrate the robustness of the CG angle measurements for the selected DES Y1 \rdmp clusters.  
    Large scatters happen at small ellipticity (caused by float errors) or near angle $\pm90\deg$ (which are equivalent). 
    Some CGs have the model type REX and thus 0 ellipticity and -90 $\deg$ angle in LS; they are partially caused by blending. 
    Only 0.7\% of the DES CGs do not have the counterpart in LS. The red dashed diagonal line is for reference.
    }
    \label{fig:angle_comparison}
\end{figure}

We compare the measurements of the CG angle using the DES Y3 and LS DR9 catalogues. We find the two methods produce consistent results (Figure \ref{fig:angle_comparison}).

\begin{figure}
    \includegraphics[width=\columnwidth]{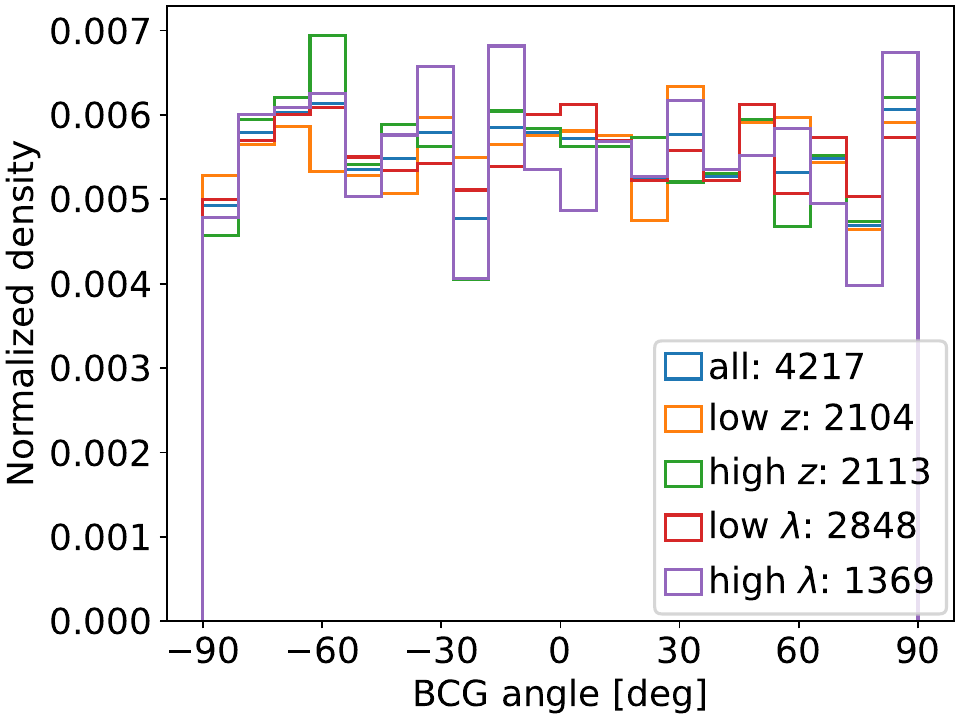}
    \caption{
    Histogram of the CG angles in different redshift/richness subsamples (with the number of clusters in each set). The histograms are normalized to show the uniformity.
    }
    \label{fig:CG_angle_hist}
\end{figure}

We also check the CG angle distributions in redshift and richness subsamples (Figure~\ref{fig:CG_angle_hist}).
The distributions  are generally uniform as expected -- we find no preferential 2D orientation of the CGs, and thus no selection bias when we make cuts on the catalogues based on the CG angle. 
Since the CG angles derived from DES and LS are consistent, we only use the DES angles in this work.

Additionally, we test and find that the difference between the CG angles of any two clusters and their angular separation does not correlate in our cluster sample; the distribution of the CG angle difference is also uniform.

\subsection{Projection of CGs and clusters}

The CG angle we measure is a projection of its 3D orientation. Though the offset between the orientation and the projection plane is likely averaged out after stacking, the detected halo ellipticity can be lowered due to the projection.
Moreover, an offset between the CG angle and the halo orientation exists~\citep{shin2018}. 

So far we have assumed the projected isodensity contours are concentric and aligned. Commonly, the contour ellipticity is assumed to be constant as well (homoeoidal), and any projection of a homoeoidal model is still homoeoidal; the projection of a varying ellipticity model produces `twisted' contours~\citep[e.g.][and references therein]{schramm1990}.  Since we stack the clusters, the twisting is likely averaged out, but the ellipticity can still vary with the radial distance (and gradually decreases to zero at sufficiently large radii).

In this paper we test two directions -- the major- and minor-axis directions of CG. Other angles, e.g. $45\deg$, from the major and minor axes can also be studied to check the symmetry, but this is beyond the scope of this paper -- we leave it to future work.


\subsection{Correlation between lensing signature and number density: Mass-number ratio} 

\begin{figure}
    \centering
    \includegraphics[width=\columnwidth]{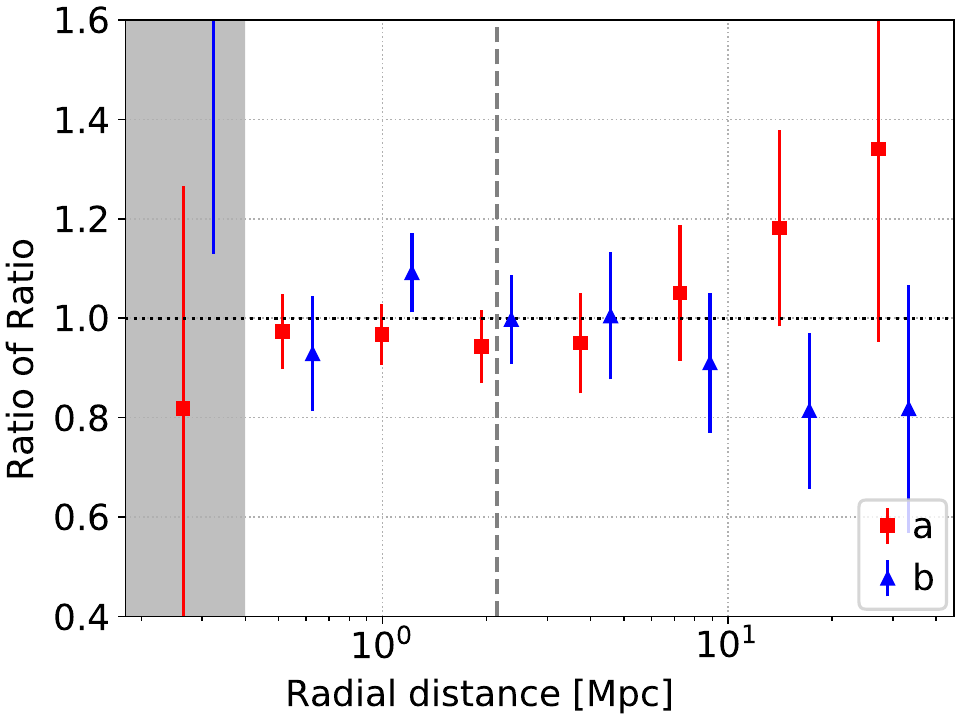}
    \caption{Ratio between the \rr1{normalized axis-aligned profiles of} effective excess surface mass density and the excess number density. The legend shows the two samples along the major axis ($a$) and minor axis ($b$) respectively.  The two ratios are nearly unity at small radii. At large radii, they diverge because the lensing signal continues to exhibit difference between the major and minor axes, and the number densities appear to be more circularized.
    }
    \label{fig:mass_number_ratio}
\end{figure}

One interesting perspective from our results is the confirmation of the simultaneous influence of cluster triaxiality on both the cluster galaxy distribution and cluster weak lensing signal, even at a very large radius ($\sim10-20$~Mpc); on average when a cluster's weak lensing signal is measured to be higher in one direction, its galaxy density is also higher in that direction. 
Similarly, a few previous studies have studied the correlation between the lensing signals and galaxy overdensities using simulations, and they find this type of correlation creates a selection effect -- clusters identified in an optical catalogue using a galaxy density criterion is likely to have a biased lensing signal. 
But we need to point out that, in previous studies, the lensing signals and galaxy overdensities are usually measured by the average over all position angles on the plane of sky (PoS), while we study them along the cluster's projected major and minor axes (traced by CG) on PoS only. 
For example, \cite{zzhang2023} studied the correlation between the clusters' richness (a probabilistic red-sequence galaxy count) and their projected radial mass densities of triaxial dark matter haloes in simulations. They concluded that triaxial haloes, when their major axis is aligned with line of sight, have both a higher richness estimation -- which is a galaxy overdensity measurement -- and a boosted lensing signal than expected from their masses. 
Our analysis provides additional \textit{observational} evidence to their finding; we find that along the major (or minor) axis of a galaxy cluster, both the cluster's galaxy overdensities and mass overdensities -- as measured by weak lensing -- will appear to be higher (or lower), and this simultaneous increase (or decrease) extends to at least $\sim10-20$~Mpc on average.

More interestingly, we show how the cluster's weak lensing signals are correlated with the galaxy overdensity measurements, depending on the radius range. In previous sections, while the cluster's lensing signals are higher along the major axis throughout the whole 0.4 to 40 Mpc radial range by $\sim20\%$, the elevation in the galaxy overdensity along the same axis does not stay elevated as far. Furthermore, along the major axis, the \textit{red-sequence} galaxy overdensity is elevated by $\sim20\%$ only to $\sim20$~Mpc, while the overdensity of \textit{non-red-sequence} galaxies stays elevated at the same level only to $\lesssim10$~Mpc. This radial-dependent change in the lensing signals and galaxy overdensities indicates that variations in the two observables are more correlated at small scales, but they may become uncorrelated at large scales. 
\rr1{We can conclude that the lensing signal is more strongly affected by halo orientation and triaxial mass distribution (to a further distance) than the observed richness. This finding is corroborated by simulation studies of  e.g.~\citet{wu2022} and~\citet{zzhang2023}. 
}

This can be further illustrated in Figure~\ref{fig:mass_number_ratio}, which compares the lensing/effective excess surface density ($\widetilde{\Delta\Sigma}$) versus photometry/excess number density ($\Delta n$), and shows the ratio of the previous two ratios, i.e., $[a/f]_{\widetilde{\Delta\Sigma}}/[a/f]_{\Delta n}$ or $[b/f]_{\widetilde{\Delta\Sigma}}/[b/f]_{\Delta n}$ (derived from the two plots in Figure~\ref{fig:canonical_dataset_ratios}). This new ratio is nearly unity at small radii, indicating an almost perfectly-correlated increase/decrease in both lensing signals and galaxy overdensities along each axis,  but the ratio gradually diverges after the effective splashback radius (especially after $\sim10$~Mpc), indicating  a weakening in the correlation. There are also hints that this ratio is even smaller (for the $a$ sample) or larger (for the $b$ sample) than 1 near the cluster centre, indicating a different degree of correlation towards the cluster core.

In galaxy cluster observations, the correlations between lensing signals and galaxy overdensities can be worsened by projections along the line of sight (LoS). When a triaxial halo is observed, if we consider the azimuthal average over all position angles on PoS, its lensing signals may appear to be stronger or weaker depending on whether its major or minor axis is aligned along LoS, and so is its observed galaxy overdensity. 
In our analysis, we study the signals along the clusters' major and minor axes on PoS, which is perpendicular to LoS, using the CG angle. 
If the same clusters are rotated to have their major axes nearly aligned with LoS, we would still expect their lensing signals to remain higher than the average of the results of all rotation angles (or the result of a spherically-symmetrical galaxy cluster with the same mass), although now this is caused by the column/surface density being higher (under the thin lens approximation) and thus not necessarily still by $\sim20\%$. Also, we expect \rr1{that} this boost is caused by the projection of not only the cluster triaxial halo \textit{but also} the nearby aligned LSS.  This boost can extend to large radii on PoS when the major axis is slightly tilted from LoS, since we find the connection between the triaxial halo and nearby filaments can extend to $\sim40$~Mpc. The case will be similar for galaxy overdensities, but we expect the boost will extend to smaller radii, \rr1{because} the alignment between the triaxial cluster galaxy distribution and nearby LSS extend to \rr1{only} $\sim10-20$~Mpc \rr1{in our results}. 
Similar arguments can be made if the same clusters in our analysis are rotated to have their minor axes aligned with LoS.  In other words, when considering projection along LoS, the clusters' observed lensing signals and galaxy overdensities may not be 1:1 correlated, and the lensing signals are likely to be more biased than the galaxy overdensities at very large radii. 

Moving forward, in recent years, the correlation between lensing signals and galaxy overdensities and its induced selection effect \rr1{has} become a crucial element of cluster cosmology analysis, 
and the selection effect has been extensively studied in simulations. 
However, our analysis is the first to provide insights into this correlation effect using observational data towards $\gtrsim20$~Mpc. As simulation studies of the correlations are often limited by their accuracy to match observational data, our method can provide crucial checks in the simulation-based conclusions.  
Further, such a correlation may exist between a variety of cluster observations, including X-ray/SZ versus lensing/richness, etc. Our analysis provides a new perspective to put observation-based constraints on this correlation effect.
\rr1{Moreover, our method may also help to provide insights into LSS galaxy bias.}

\section{Summary and Prospects}
\label{sec:summary}

In this work, we have studied galaxy cluster triaxiality by using a sub-sample of DES Y1 \rdmp clusters that have well-defined CGs. \rr1{We use the CG major and minor axes as proxies for the underlying galaxy cluster triaxial mass distribution.}
We \rr1{use these axes to build axis-aligned} (stacked) profiles of (effective) excess surface mass density ($\widetilde{\Delta\Sigma}$) via lensing, and excess galaxy number density ($\Delta n$) via photometry. \rr1{We found the following in our measurements:}
\begin{itemize}
\item \rr1{There is a clear difference between each} axis-aligned profile in both lensing signatures and galaxy distributions, especially after \rr1{normalizing} the profiles by \rr1{the} azimuthally-averaged stacked profile (as the fiducial) that considers the galaxies at all position angles. \item The normalized profiles along the two directions are nearly symmetric with respect to unity -- an increase along the major-axis direction for both lensing signatures and galaxy distributions, and decrease along the minor-axis direction. \item The mean difference \rr1{between each axis-aligned profile and the fiducial profile} is $\sim\pm10-20\%$ for both lensing signals and galaxy overdensities, starting from the inner region of the clusters ($\sim0.4$~Mpc) to nearby groups/clusters and filaments/LSS ($\sim10-20$~Mpc). 
\item \rr1{The difference between the axis-aligned profiles and the fiducial profiles} extends further out in dark matter than galaxies (the ratio of their physical lengths is $\sim2$). 
\item Since the CG orientation approximately traces the cluster's halo orientation, our results shows that nearly relaxed clusters still have clear signs of triaxiality \rr1{in the axis-aligned profiles}, which is consistent with previous studies.
\item We did not see a strong dependence \rr1{of the axis-aligned profile behavior on the cluster mass or on the cluster redshift in the two richness/redshifts bins we examined. \rr1{This is similar to the result of~\citet{Nurgaliev2017}, where the gas morphology shows no significant   redshift evolution.}}  However, the effective splashback radius ($\sim2-4$~Mpc) -- the location of a local minimum of the \rr1{fiducial} profile's derivative in the logarithmic space, does increase as the richness increases (i.e. a \rr1{more massive} cluster), as expected. 
\item The normalized mean lensing profile shows a decrease in the triaxiality signature near the effective splashback location but it increases again afterwards (a `necking' feature); while the triaxiality signature shown in the normalized mean excess galaxy number density is nearly constant and disappears  soon after $\sim10$~Mpc.
\item We also noted that the red-sequence galaxy distribution showed the same triaxiality feature but extended further out than the others, and therefore the volume they span is between that of DM and regular galaxies; the reason is likely that these red galaxies trace the matter distribution better. 
\end{itemize}
\rr1{The} primary datasets \rr1{for our analysis} are the DES Y3 and DR2 catalogues. To test the robustness of our results, we \rr1{also} used another dataset for studying the lensing signal and the galaxy number density -- the LS DR9 catalogues, and found consistency between the results of DES and LS, though they use different shape measurement and photometry methods. This indicates that our discovery does not depend on the catalogue or the measurement but is physical. 

Our method is straightforward and can easily be used on other data samples to test the cluster triaxiality. 
\rr1{Future} directions of this study \rr1{might} include using this method on other datasets from observations/simulations \rr1{(Jones et al. in preparation)}, and\rr1{/or exploring this signature of triaxiality in cluster subsamples selected} by other features. Other observational cluster samples \rr1{include} the \rdmp clusters of SDSS and DES Y6, samples in future large-area deep surveys (e.g. LSST), \rr1{and }X-ray/SZ cluster samples, which are less affected by projection. 

On the other hand, \rr1{X-ray/SZ signatures can also help determine} halo orientations~\citep{mantz2015spa,donahue16}, \rr1{motivating studies that connect} the shape of the gas distribution with that of the halo and galaxy distribution. High resolution X-ray/SZ maps provided by e.g. \textit{eROSITA} and MUSTANG will improve the measurements of \rr1{gas isocontours} and \rr1{alternatives} of the cluster centre~\citep{zhang2019_centering,bleem_2020}.
\rr1{X-ray/SZ signatures therefore enable tests of} our method on more perturbed or merging clusters (e.g. those with \rdmp $\max\{\texttt{P\_CEN}\}<0.8$), where determining the CG is difficult. 

\rr1{For alternative cluster subsample studies, one can} divide the cluster sample by the stellar masses of CGs~\citep{zu2021}, or by the gas masses to study \rr1{how these features relate to triaxial signatures in axis-aligned profiles.}
Finally, \rr1{we can apply }our method to the spectroscopic survey data (e.g. DESI) \rr1{to analyse} anisotropy in the dynamical mass distribution of a cluster sample \rr1{(Fu et al. in preparation)}. \rr1{Such an analysis would enable more detailed comparisons between} the dynamical and gravitational mass distributions of clusters.


\section*{Acknowledgements}

We thank the kind comments from the \rr1{anonymous  referee}. We are grateful for the comments and \rr1{support} from Ian Dell'Antonio, Anja von der Linden, Tae-hyeon Shin, Tesla Jeltema, Stella Seitz, C\'eline Combet, \rr1{Humna Awan, Radhakrishnan Srinivasan, LSST Dark Energy Science Collaboration (DESC) Clusters (CL) Analysis Working Group,} Tom Matheson, Aaron Meisner, Joan Najita, Arjun Dey, Adam Bolton, Abhijit Saha, \rr1{Shuang Liang, Zhuowen Zhang, Erwin T. Lau, Hyejeon Cho, Adam Wright, Eve Kovacs, Deric Jones}.

The work of SF and YZ is supported by NOIRLab, which is managed by the Association of Universities for Research in Astronomy (AURA) under a cooperative agreement with the National Science Foundation.
CA acknowledges funding from DOE grant DE-SC009193.

This research was conducted using computational resources and services at the Center for Computation and Visualization (CCV), Brown University.

This project used public archival data from the Dark Energy Survey (DES). Funding for the DES Projects has been provided by the U.S. Department of Energy, the U.S. National Science Foundation, the Ministry of Science and Education of Spain, the Science and Technology Facilities Council of the United Kingdom, the Higher Education Funding Council for England, the National Center for Supercomputing Applications at the University of Illinois at Urbana-Champaign, the Kavli Institute of Cosmological Physics at the University of Chicago, the Center for Cosmology and Astro-Particle Physics at the Ohio State University, the Mitchell Institute for Fundamental Physics and Astronomy at Texas A\&M University, Financiadora de Estudos e Projetos, Funda{\c c}{\~a}o Carlos Chagas Filho de Amparo {\`a} Pesquisa do Estado do Rio de Janeiro, Conselho Nacional de Desenvolvimento Cient{\'i}fico e Tecnol{\'o}gico and the Minist{\'e}rio da Ci{\^e}ncia, Tecnologia e Inova{\c c}{\~a}o, the Deutsche Forschungsgemeinschaft, and the Collaborating Institutions in the Dark Energy Survey.
The Collaborating Institutions are Argonne National Laboratory, the University of California at Santa Cruz, the University of Cambridge, Centro de Investigaciones Energ{\'e}ticas, Medioambientales y Tecnol{\'o}gicas-Madrid, the University of Chicago, University College London, the DES-Brazil Consortium, the University of Edinburgh, the Eidgen{\"o}ssische Technische Hochschule (ETH) Z{\"u}rich,  Fermi National Accelerator Laboratory, the University of Illinois at Urbana-Champaign, the Institut de Ci{\`e}ncies de l'Espai (IEEC/CSIC), the Institut de F{\'i}sica d'Altes Energies, Lawrence Berkeley National Laboratory, the Ludwig-Maximilians Universit{\"a}t M{\"u}nchen and the associated Excellence Cluster Universe, the University of Michigan, the National Optical Astronomy Observatory, the University of Nottingham, The Ohio State University, the OzDES Membership Consortium, the University of Pennsylvania, the University of Portsmouth, SLAC National Accelerator Laboratory, Stanford University, the University of Sussex, and Texas A\&M University.
Based on observations at Cerro Tololo Inter-American Observatory, a programme of NOIRLab (NOIRLab Prop. 2012B-0001; PI J. Frieman), which is managed by the Association of Universities for Research in Astronomy (AURA) under a cooperative agreement with the National Science Foundation.

The Legacy Surveys consist of three individual and complementary projects: the Dark Energy Camera Legacy Survey (DECaLS; Proposal ID \#2014B-0404; PIs: David Schlegel and Arjun Dey), the Beijing-Arizona Sky Survey (BASS; NOAO Prop. ID \#2015A-0801; PIs: Zhou Xu and Xiaohui Fan), and the Mayall z-band Legacy Survey (MzLS; Prop. ID \#2016A-0453; PI: Arjun Dey). DECaLS, BASS and MzLS together include data obtained, respectively, at the Blanco telescope, Cerro Tololo Inter-American Observatory, NSF's NOIRLab; the Bok telescope, Steward Observatory, University of Arizona; and the Mayall telescope, Kitt Peak National Observatory, NOIRLab. The Legacy Surveys project is honored to be permitted to conduct astronomical research on Iolkam Du'ag (Kitt Peak), a mountain with particular significance to the Tohono O'odham Nation.
BASS is a key project of the Telescope Access Programme (TAP), which has been funded by the National Astronomical Observatories of China, the Chinese Academy of Sciences (the Strategic Priority Research Programme `The Emergence of Cosmological Structures' Grant \# XDB09000000), and the Special Fund for Astronomy from the Ministry of Finance. The BASS is also supported by the External Cooperation Programme of Chinese Academy of Sciences (Grant \# 114A11KYSB20160057), and Chinese National Natural Science Foundation (Grant \# 11433005).
The Legacy Survey team makes use of data products from the Near-Earth Object Wide-field Infrared Survey Explorer (\textit{NEOWISE}), which is a project of the Jet Propulsion Laboratory/California Institute of Technology. \textit{NEOWISE} is funded by the National Aeronautics and Space Administration.
The Legacy Surveys imaging of the DESI footprint is supported by the Director, Office of Science, Office of High Energy Physics of the U.S. Department of Energy under Contract No. DE-AC02-05CH1123, by the National Energy Research Scientific Computing Center, a DOE Office of Science User Facility under the same contract; and by the U.S. National Science Foundation, Division of Astronomical Sciences under Contract No. AST-0950945 to NOAO.
The Photometric Redshifts for the Legacy Surveys (PRLS) catalogue used in this paper was produced thanks to funding from the U.S. Department of Energy Office of Science, Office of High Energy Physics via grant DE-SC0007914.

This research uses services or data provided by the Astro Data Lab at NSF’s NOIRLab. NOIRLab is operated by the Association of Universities for Research in Astronomy (AURA), Inc. under a cooperative agreement with the National Science Foundation.

In this work we use these softwares:
Numpy~\citep{Harris2020}, 
Scipy~\citep{Virtanen2020},
Astropy~\citep{Astropy2013,Astropy2018},
Matplotlib~\citep{Hunter2007}, 
Healpy~\citep{2005ApJ...622..759G,Zonca2019},
H5py\footnote{\url{https://github.com/h5py/h5py}},
IPython~\citep{Perez2007}.
Some of the results in this paper have been derived using the healpy and HEALPix packages.


\section*{Data Availability}

The DES and DESI Legacy Image Survey catalogues described in this paper can be obtained from the DES Data Management  website\footnote{\url{https://des.ncsa.illinois.edu}} and the NOIRLab Astro Data Lab website\footnote{\url{https://datalab.noirlab.edu}}. 
The data generated in our results are available from the corresponding author upon reasonable request.




\bibliographystyle{mnras}
\bibliography{main} 




\appendix

\section{Consistency of results using the Legacy Surveys cluster catalogue}\label{sec:ls_cluster}

We test using the Legacy Surveys cluster catalogue~\citep{zou21}, which is derived from the Legacy Surveys DR8 catalogue\footnote{\url{https://www.legacysurvey.org/dr8/}}, and using DES Y3 shapes~\citep{gatti21} for lensing analysis and DES DR2 photometry~\citep{abbott21} for number density analysis.
The Legacy Surveys cluster catalogue uses galaxy concentration and
photo-z to detect clusters without identifying the red sequence. The catalogue covers the whole $\sim14000\deg^2$ of the Legacy Surveys sky footprint in $g,r,z$-bands ($\sim20000\deg^2$ for at least one pass in one band), which includes the full $\sim5000\deg^2$ of the DES footprint. Therefore, only the clusters detected in/near the DES footprint will be analysed here, but this is still larger than the footprint of the DES Y1 \rdmp cluster catalogue which covers $\sim1500\deg^2$.

\begin{figure*}
    \centering
    \includegraphics[width=\columnwidth]{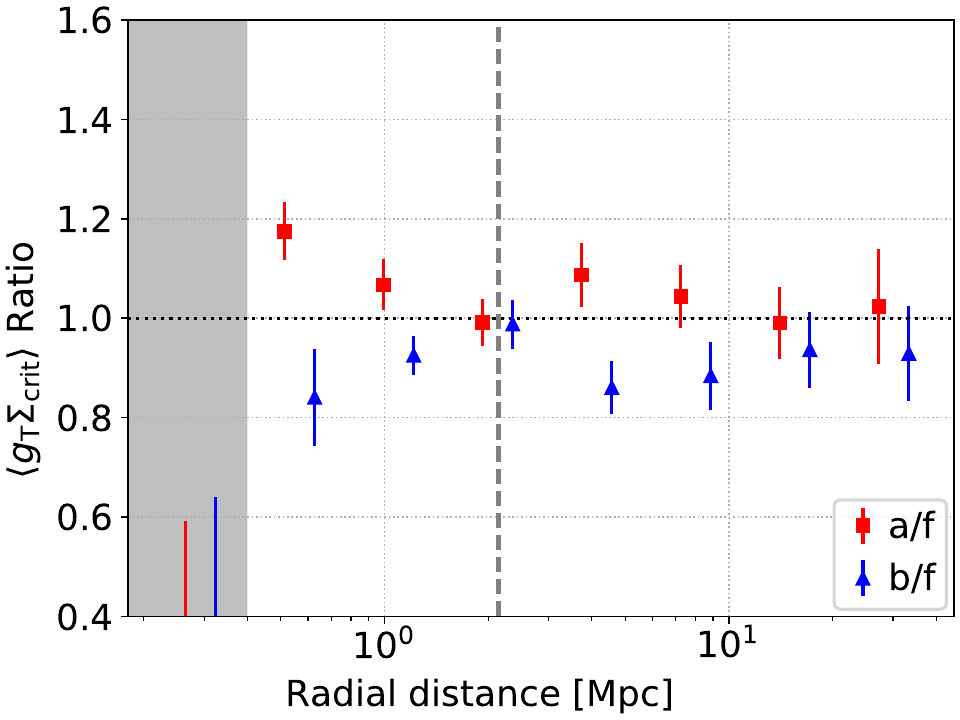}
    \includegraphics[width=\columnwidth]{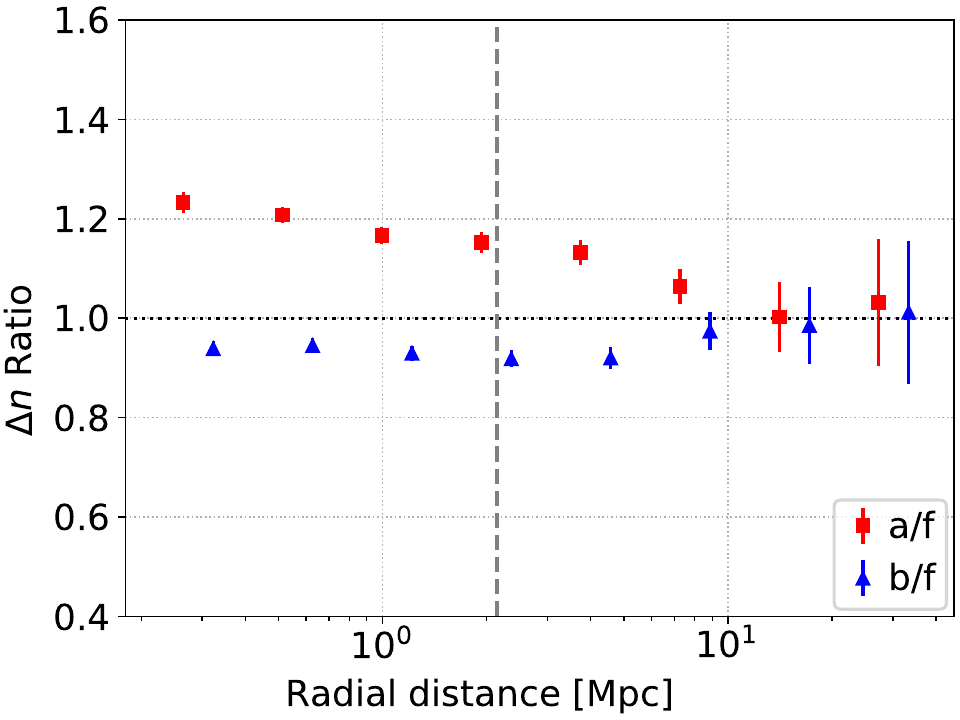}
    \caption{
    \rr1{Normalized axis-aligned }lensing profiles (\textit{left}) and excess number density profiles (\textit{right}) using a sample of LS clusters that are nearly relaxed.  
    The shapes and photometry are still from DES Y3 and DES DR2.
    The lensing analysis includes 13147 clusters for the fiducial and 11813 clusters for the $a, b$ subsamples, while the number density analysis includes 13310 and 11817 clusters respectively. 
    }
    \label{fig:LS_cluster}
\end{figure*}

We use the method described in Section \ref{sec:cluster_sample} to select a sample of Legacy Surveys clusters which are nearly relaxed and well detected. We use the same method mentioned in Section \ref{sec:shape_catalog} and \ref{sec:photometry_catalog} to build cluster profiles and stack them, except when we make individual number density profiles, we set the empty bins to NaN, to reduce the effect of footprint boundaries. 
We present the results in Figure \ref{fig:LS_cluster}.

We find that the general trend of the ratios are consistent with the previous ones derived from the DES Y1 \rdmp clusters (Figure~\ref{fig:canonical_dataset_ratios}), but the signatures of triaxiality  are slightly smaller here. We still can see the `necking' features, but the lensing signal ratio seems to converge at large radii. 
The reason could be that the ratios are sensitive to how the cluster central galaxies are selected, and to the dynamical states of the clusters; we lack a metric to well determine them here. 
When we cross match our Legacy Surveys cluster sample and our DES Y1 \rdmp cluster sample, and compare the distance between the galaxy number density peak and BCG versus $\max\{ \texttt{P\_CEN}\}$ in the respective catalogues, we find
concentrations of clusters that have the distance $\sim0$~Mpc or $\max\{ \texttt{P\_CEN}\}\sim1$, but there are also clusters where the two quantities do not appear to be correlated.

\section{Individual richness/redshift bins for excess number density ratio}\label{sec:individual_bin}

\begin{figure*}
    \includegraphics[width=1.95\columnwidth]{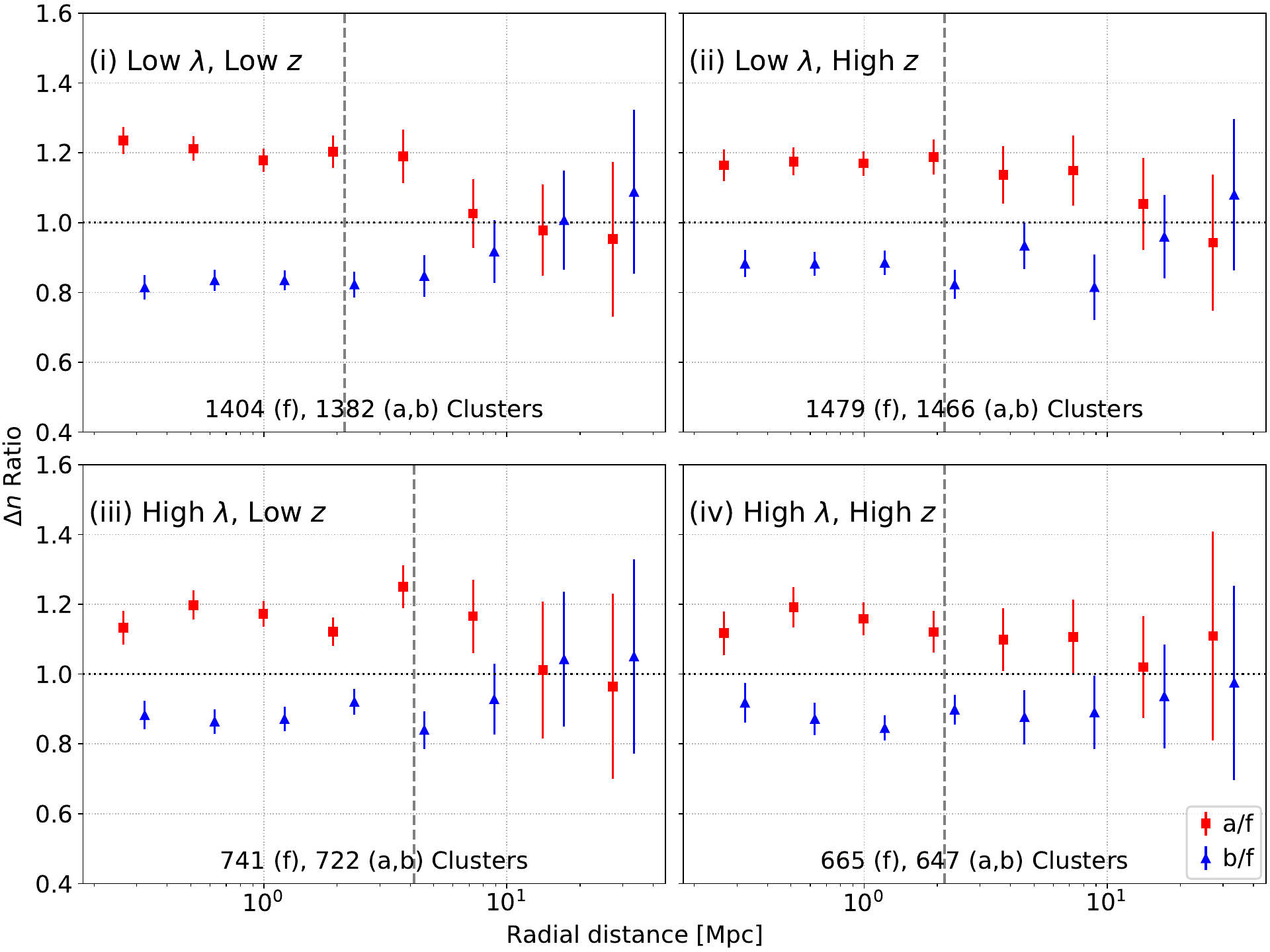}
    \caption{\rr1{Normalized axis-aligned excess number density profiles }in individual richness ($\lambda$) and redshift ($z$) bins. Note we decrease the jackknife region numbers to 72 and 64 for the high-$\lambda$ low-z and high-$\lambda$ high-z datasets respectively because of the limited number of clusters (others are using the default 100 regions). The low-richness low-redshift bin shows the smallest converging radius.
    }
    \label{fig:number_individual}
\end{figure*}

We divide the cluster sample into four subsamples by both redshift and richness, which both have two bins, and stack the excess galaxy number density in each bin to compute respective normalized profiles (Figure \ref{fig:number_individual}). The result is much noisier than the one in the main text, which uses two bins for redshift or richness only (Figure \ref{fig:bins_number}; Section \ref{sec:richness_redshift}). 
We test the same $2\times2$ binning for the lensing data and find the result is even noiser. 
Therefore, in the main text we only use two bins rather than four bins to divide the cluster sample by redshift or richness.

\section{Flow chart for the Pipeline}\label{sec:flow_chart}

\begin{figure*}
    \includegraphics[width=1.95\columnwidth]{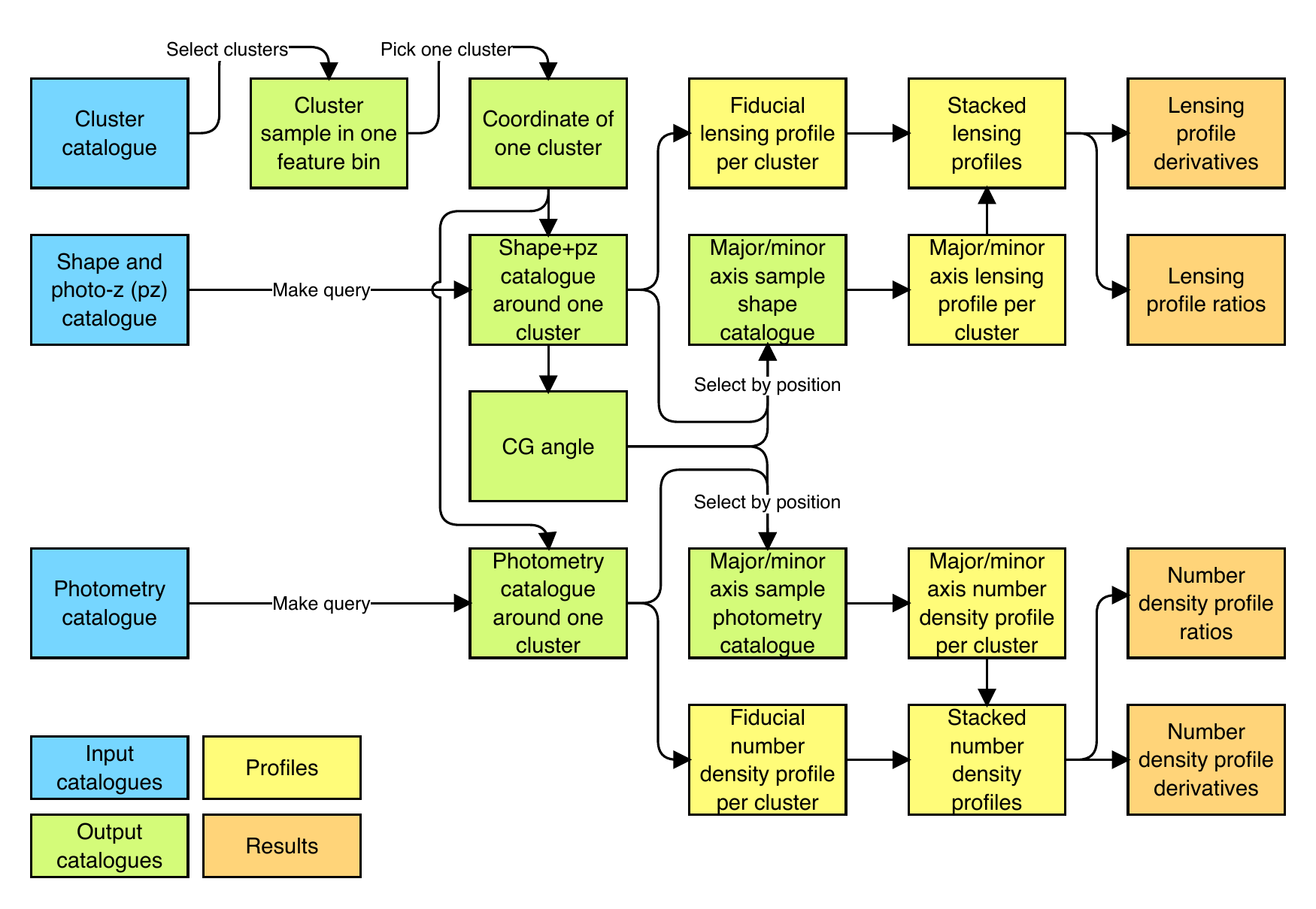}
    
    \caption{Flow chart of our pipeline. We note that obtaining a CG angle do not require photo-z information, and thus CGs can be queried independently using small cone searches. Further cuts can be made on the photometry catalogues, e.g. a red-sequence cut (Section \ref{sec:red-sequence}).
    }
    \label{fig:flow_chart}
\end{figure*}

The flow chart (Figure \ref{fig:flow_chart}) illustrates our pipeline -- it starts from selecting a cluster sample and making queries for catalogues, then to computing the CG angle and selecting sources based on the angle, and to stacking the catalogues, and finally to producing the stacked profiles and their ratios.

\bsp	
\label{lastpage}
\end{document}